\documentclass[journal]{IEEEtran}
\usepackage{xcolor}
\usepackage{multirow}
\usepackage{todonotes}
\usepackage{float}
\usepackage{subcaption}
\usepackage{comment}
\usepackage{svg}
\usepackage{url}
\usepackage{booktabs}
\usepackage{amsmath}
\newcommand{\parhead}[1]{\paragraph{#1}\mbox{}}

\AtBeginDocument{%
  }

\begin{document}

\title{Semantic Validation of Packer Identification Tools: Characterization, Repair, and Downstream Impact}


\author{Fangtian~Zhong,~\IEEEmembership{Member,~IEEE,}
			 Zhuoyun Qian,
			Yili Jiang,~\IEEEmembership{Member,~IEEE,}
            Jiaqi Huang,~\IEEEmembership{Member,~IEEE,}
           Yunming Pang,
           Xiuzhen Cheng,~\IEEEmembership{Fellow,~IEEE}
			\thanks{Fangtian~Zhong, Zhuoyun Qian and Xiuzhen Cheng are with College of Computer Science and Technology, Shandong University, Qingdao,
				Shandong 266237, China. E-mail: fzhong@sdu.edu.cn, 202699900021@email.sdu.edu.cn, xzcheng@sdu.edu.cn}
			\thanks{Yili~Jiang and Jiaqi~Huang are with  Department of Computer Science, Georgia State University, Atlanta, GA 30303, USA. E-mail: \{yjiang27, jhuang45\}@gsu.edu}
            \thanks{Yunming Pang is with Department of Computer Science, Northeast Normal University, Changchun, Jilin 130000, China. E-mail: pangym527@nenu.edu.cn}
        }




\maketitle
\begin{abstract}
Packer identification tools are essential to malware analysis, yet their semantic correctness remains largely unverified. Incorrect packer labels can trigger inappropriate unpacking and degrade downstream malware classification. This paper presents a framework for systematically exposing, characterizing, and repairing semantic errors in packer identification tools. Our key insight is to leverage unpackers as executable semantic contracts, enabling automated validation of packer predictions without requiring manually labeled packer ground truth. We develop a pipeline that normalizes heterogeneous tool outputs, localizes faulty decision logic, and repairs identification errors through complementary signatures, heuristic rules, and static unpacker-derived detection predicates. Our empirical evaluation reveals widespread identification failures, with an average packer family identification recall of only 44.8\% and an average packedness detection recall of only 63.6\%. The proposed two-stage repairs achieve an approximately 100.2\% relative improvement in average packer family identification recall, while increasing packedness detection recall to 99.0\%-99.5\%. We further demonstrate that semantic errors can propagate through unpacking failures and adversely affect downstream malware classification. Correcting these errors improves average classification precision by 13.6 percentage points.

\end{abstract}
\begin{IEEEkeywords}
Packer, Semantic Errors, Malware Classification
\end{IEEEkeywords}
\section{Introduction}

Malware authors widely use packers to obstruct static analysis and increase the cost of reverse engineering. By compressing, encrypting, or transforming executable contents, packers produce syntactically altered binaries that conceal the original program content \cite{stephens2018probabilistic}. Before analysts can reliably inspect such binaries or determine their malware families, the packed content must first be recognized and, when possible, unpacked. Consequently, packer identification has become a foundational component of modern malware analysis pipelines, directly affecting downstream tasks such as unpacking \cite{zheng2025gupacker,zhong2025unveiling} and malware classification \cite{zhong2024enhancing,li2025malmixer}. Despite this central role, the outputs of packer identification tools are typically treated as trustworthy by default and are often used as de facto ground truth by both researchers and practitioners \cite{downing2021deepreflect,maleki2019improved,bhardwaj2021reverse}.

However, the reliability of these predictions remains insufficiently validated. Existing packer identification tools employ diverse detection strategies, including signature matching, entropy-based heuristics, graph matching, and machine learning. Although practical and widely used, these approaches generally rely on indirect syntactic, structural, or statistical evidence to determine whether an executable is packed or to identify its packer family. Such evidence may be incomplete, ambiguous, or shared across different packer families, leading to missed detections or incorrect packer labels. These errors are particularly consequential because packer identification frequently guides unpacker selection \cite{zhong2025unveiling}. A missed detection may cause the analysis pipeline to skip unpacking, while an incorrect packer label may trigger an inappropriate unpacker. In either case, the original program content may remain inaccessible, propagating identification errors into subsequent unpacking and malware analysis stages.

Unfortunately, systematically validating the semantic correctness of packer identification tools is difficult. First, reliable ground-truth labels for real-world packed binaries are often unavailable, making it hard to determine which tool output is actually correct. Second, different tools expose highly heterogeneous outputs, ranging from packedness decisions to ranked file-type descriptions or specific packer family labels. Moreover, even after output normalization, disagreements remain difficult to interpret because different tools are built on fundamentally different decision logics. These challenges make semantic faults hard to detect, explain, and repair.

In this paper, we address these challenges by introducing a semantic error exposure framework for packer identification tools. Our key idea is to leverage unpackers as executable semantic contracts, where successful unpacking provides evidence that a predicted packer label is consistent with the underlying packing mechanism. Based on this insight, we construct test oracles from unpacking outcomes to expose semantic inconsistencies without requiring manually labeled ground truth. We further develop a systematic pipeline that normalizes heterogeneous tool outputs, localizes faulty decision logic, and repairs identification errors through complementary signatures, heuristic rules, and static detection predicates derived from unpacker implementations. Compared with identifying packers through dynamic unpacking, our approach avoids executing samples and reconstructing their contents during identification, reducing the processing required for each input. Its family predictions also narrow the set of candidate unpackers, reducing the need for repeated unpacking attempts and supporting large-scale malware screening.

Using this framework, we conduct a large-scale empirical study of semantic faults across publicly available packer identification tools and proprietary tools available through VirusTotal. Our findings reveal widespread identification failures stemming from incomplete signatures and brittle, family-dependent heuristics. We further demonstrate that these failures extend beyond packer identification, as incorrect packer labels can prevent successful unpacking and preserve packing-induced distortions in the features used for malware classification. Our targeted repairs substantially improve packer identification recall and downstream classification performance. In summary, this paper makes the following contributions:

\begin{itemize}
    \item We propose a semantic error exposure framework that leverages unpackers as executable semantic contracts, enabling automated validation of packer identification results and exposure of semantic inconsistencies without manually labeled ground truth.
    
    \item  We conduct the first large-scale cross-tool empirical study that characterizes recurring identification failures and reveals limitations of existing signature-based and heuristic-based detection strategies.
    
    \item We develop a two-stage repair framework that integrates complementary signatures, heuristic rules, and static unpacker-derived detection predicates, achieving an approximately 100.2\% relative improvement in packer family identification recall and increasing packedness detection recall to 99.0\%--99.5\%.
    
    \item We demonstrate the downstream impact of packer identification errors on malware classification and show that correcting these errors improves average classification precision by 13.6 percentage points, highlighting the importance of semantic correctness in malware analysis pipelines.
\end{itemize}

\section{Related Work and Motivation}
\label{sec:related}
We first introduce the related work in this section and then motivate our work through a concrete example that highlights the challenges of testing packer identification tools.
\subsection{Related Works} 
Packer detection techniques in existing works primarily rely on signatures, heuristics,  machine learning or graph matching.

\noindent\textit{\textbf{Signature-based Methods}} A major line of prior work identifies packers through signatures, based on the observation that known packers may leave recognizable artifacts such as byte patterns, instruction sequences, section names, and string markers. Tools including app-peid \cite{apppeid_github}, PEiD \cite{peid_github}, PyPackerDetect \cite{pypackerdetect_github}, and readpe \cite{readpe_github} compare such artifacts against packer signatures maintained in their own databases or drawn from public collections, including ASL \cite{asl_github}, MalScan \cite{malscan_github}, and PEiD Tab \cite{peid_top4download}. Related tools, such as qu1cksc0pe \cite{qu1cksc0pe_github}, pypeid \cite{ffri_pypeid}, and Manalyze \cite{manalyze_github}, also use YARA rules or packer-specific markers. TrID \cite{trid_website} follows a related definition-based approach, using probabilistic matching to rank candidate packers.

\noindent\textit{\textbf{Heuristic-based Methods}} A complementary line of work studies packedness detection, i.e., deciding whether an executable is packed based on generic structural or statistical symptoms rather than identifying a specific packer. Early approaches mainly rely on entropy-based and entry-point-region heuristics. REMINDer \cite{han2009packed} checks whether the entry-point section is writable and sufficiently high in entropy. Bintropy \cite{lyda2007using} uses block-wise or file-wise Shannon entropy and classifies a sample as packed when both average and maximum block entropy exceed predefined thresholds. pypeid \cite{ffri_pypeid} follows the same entropy-based intuition and labels a valid PE file as packed when its Shannon entropy exceeds a threshold. Other tools combine multiple lightweight heuristic rules. PyPackerDetect \cite{pypackerdetect_github}, Manalyze \cite{manalyze_github}, readpe \cite{readpe_github}, and Qu1cksc0pe \cite{qu1cksc0pe_github} use different combinations of rules such as suspicious section names, anomalous entry-point placement, executable-and-writable or high-entropy sections, import tables, overlay data, modified DOS stubs, and abnormal section boundaries.

\noindent\textit{\textbf{Other Methods}} Other work explores hybrid, machine learning-based, and graph-based packer identification. Detect-It-Easy (DIE) \cite{die_github} combines signature matching with lightweight heuristic rules, such as imports, sections, and API presence. Machine learning methods instead formulate packer classification as supervised classification over engineered representations, including static PE features \cite{biondi2019effective}, binary visualizations such as Byte plots and Markov plots \cite{kancherla2016packer}, and randomness profiles extracted from sliding-window analysis \cite{sun2010pattern}. Other representations from dynamic analysis aim to reduce sensitivity to static obfuscation by using runtime evidence, such as system-call traces collected in a sandbox \cite{zhang2018sensitive} or entropy sequences observed during unpacking execution \cite{bat2017entropy}. Graph-based approaches exploit structural representations of packed binaries for packer identification. Hai \emph{et al.} \cite{hai2017packer} construct unpacking-code CFGs and summarize obfuscation metadata as packer features. Saleh \emph{et al.} \cite{saleh2017control} derive structural features from entry-point CFGs, while Li \emph{et al.} \cite{li2019consistently} compare binaries using consistently-executing graphs and graph-kernel-based similarity. Combining machine learning with graph matching, PackHero \cite{di2025packhero} employs a Graph Matching Network and clustering-based similarity aggregation to identify packers through call graph matching.

\subsection{Motivating Example}

\begin{figure}[t!]
    \centering
    \includegraphics[scale=0.45]{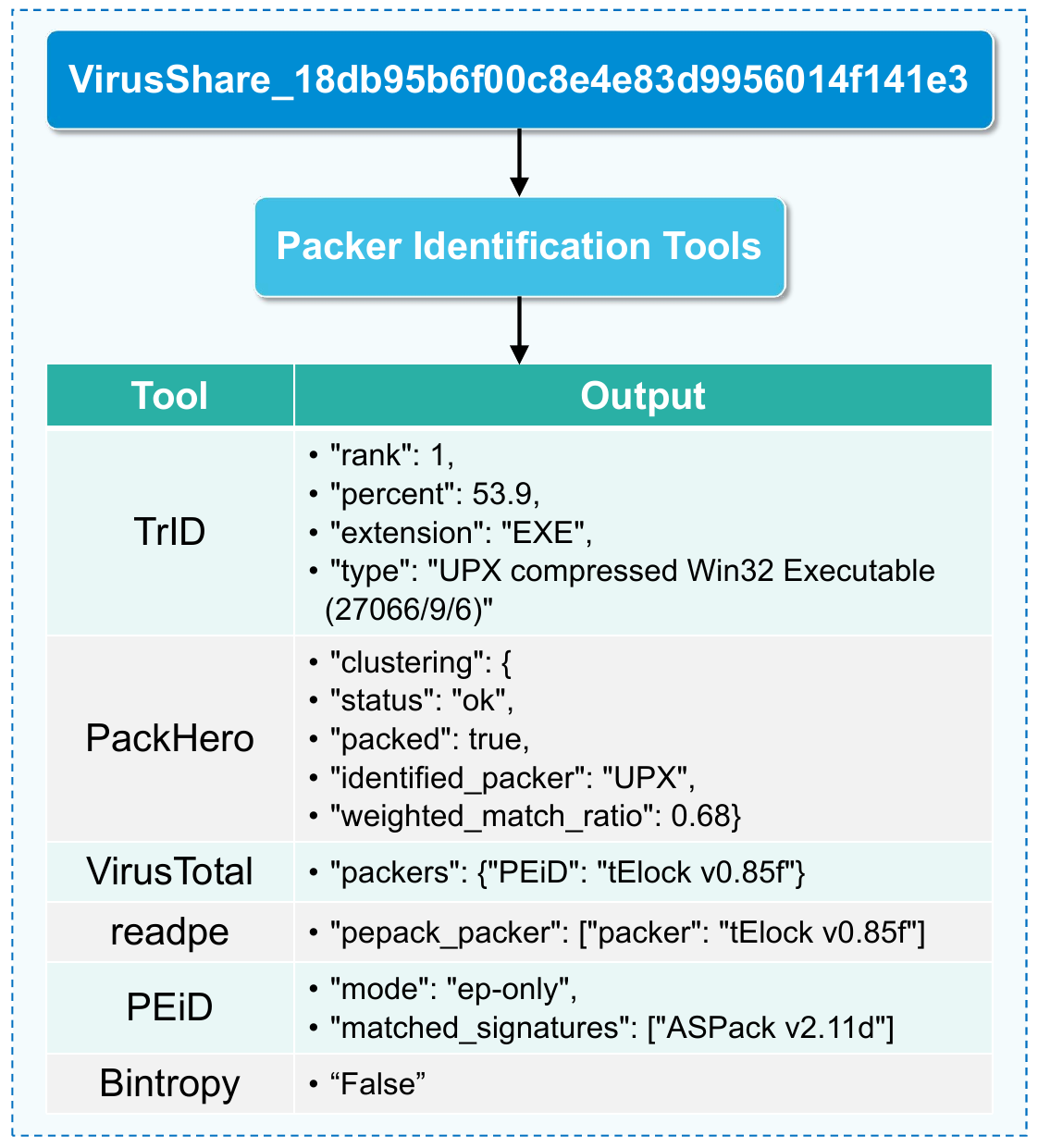}
    \captionsetup{skip=2pt}
    \caption{An output example of packer identification tools}
    \label{fig:example}
    \vspace{-5mm}
\end{figure}

We begin with a simple yet representative example involving contradictory outputs from various packer identification tools. For the binary in Fig.~\ref{fig:example}, different packer identification tools assign substantially different labels. TrID reports \texttt{UPX compressed Win32 Executable}, PackHero identifies \texttt{UPX}, PEiD matches \texttt{ASPack v2.11d}, VirusTotal and readpe report \texttt{tElock v0.85f}, while Bintropy only returns \texttt{False}. None of these results is consistent with the actual packer \texttt{Themida} \cite{themida_net_unpacker} revealed by unpacking. Therefore, this case is not merely an instance of heterogeneous output format and decision logic. Instead, it is a concrete semantic inconsistency in which different tools assign semantically incorrect packer labels to the same binary. Once the assigned label is wrong, the analysis pipeline may invoke an inappropriate unpacker, leading to failed unpacking. Since many downstream malware analysis tasks, such as static feature extraction, control-flow recovery, and behavioral inspection, rely on correctly unpacked binaries, the error can further propagate beyond packer identification and ultimately compromise the reliability of subsequent malware analysis. To systematically expose semantic faults in packer identification tools, we have to cope with the following three challenges.

\parhead{No Ground Truth} Systematically testing packer identification tools requires reliable ground-truth labels, yet obtaining them at scale is difficult. Real-world malware corpora contain massive numbers of packed samples, making manual labeling costly and often impractical. Moreover, when different tools produce conflicting labels, determining the correct one typically requires manual reverse engineering to inspect the actual packing mechanism and match it to a specific packer. This process is time-consuming and expertise-intensive.

\parhead{Heterogeneous Output Format} Besides, as shown in Fig. \ref{fig:example},
packer identification tools produce outputs with substantially different formats and levels of granularity. Some return only a binary packed/unpacked decision, while others report specific packer families, confidence scores, or ranked file-type descriptions. As a result, outputs from different tools cannot be directly compared without additional normalization. This makes it difficult to determine whether disagreements reflect genuine detection errors or simply differences in output representation.

\parhead{Heterogeneous Decision Logic} Finally, packer identification tools differ substantially in their underlying decision logic. While some rely on heuristics or signature matching, others combine both. These tools may also vary in threshold settings, heuristic design, and signature databases. Consequently, disagreements among tools often reflect deeper differences in how evidence is interpreted, making inconsistent predictions difficult to explain, faults difficult to localize, and systematic repair challenging.

\section{Semantic Validation \& Downstream Impact}
\label{sec:design}

\subsection{Overview} 

\begin{figure*}[t]
    \centering
    \includegraphics[width=0.92\linewidth]{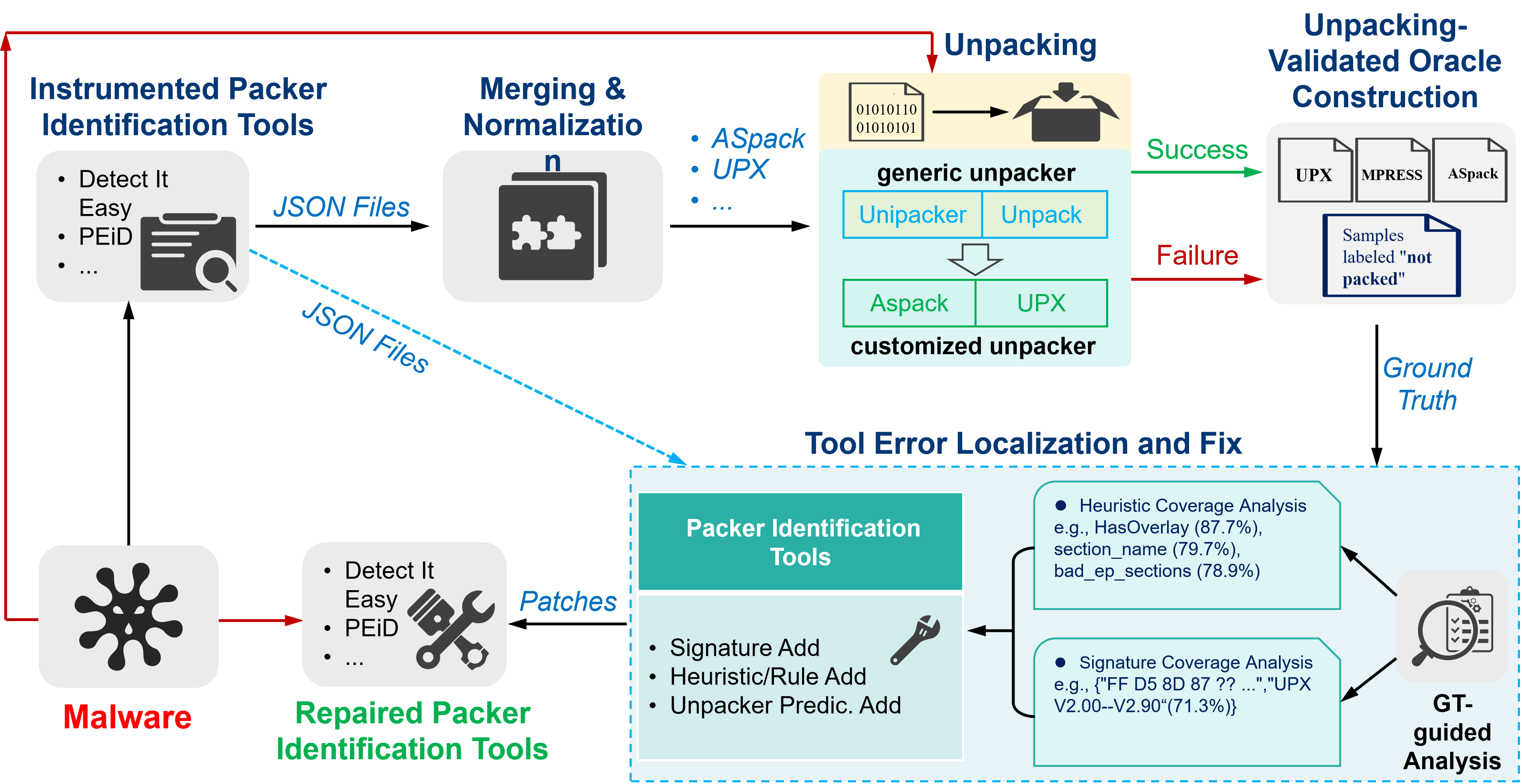}
    \captionsetup{skip=2pt}
    \caption{Overall workflow of semantic validation and static rule repair}
    \label{fig:workflow}
    \vspace{-3mm}
\end{figure*}

As illustrated in Fig.~\ref{fig:workflow}, we present a framework for systematically exposing and repairing semantic faults in packer identification tools using unpackers as executable semantic contracts. Given a benchmark of binary executables, we first collect predictions from multiple packer identification tools and apply generic and customized unpackers in a unified pipeline to construct unpacking-validated test oracles. We then instrument the packer identification tools to produce structured, traceable outputs, which are merged and normalized to enable consistent cross-tool comparison. By correlating these outputs with the unpacking-validated oracles, we localize faulty decision logic and develop targeted repairs for packer family identification and packedness detection through complementary signatures, heuristic rules, and static unpacker-derived detection predicates. This framework enables systematic fault detection, localization, and repair without requiring manually labeled ground truth.

\subsection{Establishing Test Oracles}

We collected results by executing existing packer identification tools and querying VirusTotal~\cite{virustotal} on our dataset. To establish test oracles, we integrated available generic and customized unpackers into a unified pipeline. Each binary was first processed by generic unpackers, followed by customized unpackers corresponding to the predicted packer families. For samples that could not be successfully unpacked through these steps, we applied other available unpackers as a fallback to identify a compatible unpacking procedure. Successful unpacking outcomes were then used to construct unpacking-validated packer labels, providing executable evidence of the underlying packing mechanisms. These labels serve as automatically derived test oracles for assessing the semantic consistency of packer identification results and identifying potentially faulty signatures or heuristic rules without requiring manually labeled ground truth. The unpacker collection serves as a validation resource for assessing and repairing static identification rules. Identifying packers through unpacker execution can require multiple recovery attempts when the family is unknown. Static identification provides an initial decision before these recovery operations, guiding unpacker selection without executing the sample or recovering its contents. Our framework uses the semantic evidence obtained during validation to improve this decision for subsequent analysis.

\subsection{Unifying Heterogeneous Output}

This process consists of two stages, \textit{code instrumentation} and \textit{result merging and normalization}. In the first stage, we instrument each packer identification tool to generate structured JSON outputs while preserving its original decision logic. For signature-based methods, we record the predicted packer labels and the corresponding matched signatures extracted from their signature databases. For heuristic-based methods, we record the source-code locations of their detection rules to facilitate subsequent fault localization.

In the second stage, we merge and normalize the instrumented outputs to enable systematic cross-tool comparison. Based on their detection objectives, we distinguish two evaluation targets: \textit{packedness detection} and \textit{packer family identification}. Heuristic-based decisions are normalized into binary packed/unpacked predictions, whereas signature-based, hybrid, graph-based, and learning-based identification results are normalized into packer family labels. This distinction allows us to evaluate heterogeneous detection mechanisms according to their respective objectives. We then aggregate the outputs of all tools for each executable into a unified JSON record and simplify the representation to retain the results of individual detection methods, e.g., Tool name'': \{heur'': yes/no, ``signature match'': packer family\}. The resulting representation supports consistent evaluation, fault localization, and subsequent repair across different identification tools.

\subsection{Repairing Heterogeneous Decision Logic}
We develop targeted repair strategies for two types of packer identification tasks: \textit{packer family identification} and \textit{packedness detection}. Both strategies follow a two-stage process that first integrates complementary detection evidence from existing tools and then leverages unpacker-derived detection logic to address the remaining identification failures.

For \textit{packer family identification}, the first stage integrates validated packer-specific signatures collected from different tools. We compare the original identification results with the unpacking-validated ground truth to identify signatures associated with correct packer labels. These signatures are combined into a shared family-specific signature pool to recover samples missed by individual tools. In the second stage, we examine the packer identification logic embedded in publicly available unpackers and extract additional static detection predicates. These predicates inspect structural and syntactic properties, such as PE sections, imports, entry-point patterns, and metadata, to identify packer families that remain poorly covered by existing signatures. 

For \textit{packedness detection}, the first stage augments each tool's original heuristic decision with complementary rules selected from other tools. We analyze the effectiveness of individual heuristic rules across packer families and combine rules that capture different packing artifacts. In the second stage, we examine residual false negatives and extract supplementary static detection predicates from publicly available unpackers to address the remaining detection gaps. These predicates capture packing-related evidence not adequately represented by the existing heuristics, thereby improving packedness detection coverage.

Together, these strategies combine cross-tool detection evidence with unpacker-derived rules to improve both packer family identification and packedness detection.

\subsection{Conducting Downstream Malware Analysis}

To evaluate the downstream impact of semantic errors in packer identification tools, we constructed a malware classification dataset using a labeling procedure inspired by prior work~\cite{zhong2024enhancing}. We then consulted Malpedia~\cite{Malpedia} and other publicly available malware databases to verify the assigned malware labels. We also analyzed the samples using the open-source dynamic analysis system CAPE~\cite{capev2} and performed manual reverse engineering to further validate these labels. To ensure labeling reliability, we used Cohen's Kappa coefficient~\cite{vieira2010cohen} to measure inter-author agreement. Two authors independently labeled 5\% of the malware samples, achieving a Cohen's Kappa of approximately 0.81. After a calibration discussion, they independently labeled an expanded subset comprising 10\% of the dataset, including the initial 5\%, increasing Cohen's Kappa to 0.94. The remaining samples were then labeled in nine iterative rounds, each covering an additional 10\% of the dataset. Throughout this process, Cohen's Kappa consistently remained above 0.9, and disagreements were resolved by a third author. After resolving labeling disagreements, we finalized the dataset by assigning each sample to a malware category based on its class and family labels. Following prior work~\cite{zhong2025unveiling}, we retained only categories containing more than 90 samples. The final dataset comprises approximately 27K samples spanning 46 malware categories. To assess the downstream impact of packer identification errors and their repairs, we reimplemented representative malware classification methods from prior studies~\cite{zhong2024enhancing,zhong2022malware}. We compare their classification performance before and after correcting packer identification results and applying the corresponding unpackers, allowing us to evaluate how identification errors affect subsequent malware classification.

\section{Evaluation}
\label{sec:evaluation}

Our empirical study aims to answer the following research questions:

\parhead{RQ1} To what extent do existing packer identification tools exhibit semantic errors in packer family identification and packedness detection?

\parhead{RQ2} How effective are the proposed two-stage repair strategies in addressing semantic inconsistency and improving packer identification performance?

\parhead{RQ3} How do packer identification errors and their subsequent repairs affect downstream malware classification performance?

\subsection{Experimental Setup}
\label{subsec:setup}
We evaluate our approach on eleven publicly available packer identification tools, including Bintropy~\cite{lyda2007using}, REMINDer~\cite{han2009packed}, PEiD~\cite{peid_github}, TrID~\cite{trid_website}, pypeid~\cite{ffri_pypeid}, PyPackerDetect~\cite{pypackerdetect_github}, Manalyze~\cite{manalyze_github}, readpe~\cite{readpe_github}, qu1cksc0pe~\cite{qu1cksc0pe_github}, Detect It Easy (DIE)~\cite{die_github}, and PackHero~\cite{di2025packhero}. Among these tools, TrID is proprietary, while the remaining tools are open-source. We also evaluate six proprietary packer identification tools available through VirusTotal~\cite{virustotal}, including VT PEiD, VT Cyren, VT Varist, VT F-PROT, VT Command, and VT Taggant. To ensure a broad evaluation scope, we collected publicly available packer identification tools through systematic keyword-based searches using terms such as packer identification tools, packer identifiers, and packer recognition. The resulting set includes tools that directly report packer-related labels or packedness decisions and represent diverse detection strategies. We exclude PEiD-derived variants, including ASL~\cite{asl_github}, MalScan~\cite{malscan_github}, and PEiD Tab~\cite{peid_top4download}, because their signature databases substantially overlap with that of PEiD~\cite{peid_github}. Including these variants would introduce redundant evaluations of overlapping signatures rather than substantially independent identification capabilities.

Our evaluated tools span signature-based, heuristic-based, graph-based, and machine learning-based techniques. We include the proprietary VirusTotal tools because their packer labels are externally observable and can therefore be evaluated within our framework. Our repair, however, focuses on open-source signature-based and heuristic-based tools, whose decision logic is directly accessible and modifiable. We do not repair proprietary tools because their implementations are unavailable. Graph-based and learning-based methods, such as PackHero, are also outside the scope of our repair strategies because their modification generally requires changes to structural representations, models, or training procedures rather than direct signature or heuristic augmentation. Therein, PackHero has reported the best performance among many graph-based and learning-based methods ~\cite{di2025packhero}. Moreover, learning-based methods that rely on labels generated by existing packer identification tools may inherit their semantic errors, requiring further investigation.

Our evaluation dataset consists of approximately 160K Windows PE executables randomly sampled from VirusShare~\cite{virusshare}, comprising approximately 130K samples in the initially analyzed dataset and 30K samples in a separate test dataset. The initially analyzed dataset is used to characterize identification failures and construct repair rules, whereas the separate test dataset is used to evaluate the resulting repair configurations on unseen samples. The samples cover a ten-year period from 2017 to 2026. We focus on PE executables because packer identification and unpacking are most mature and widely supported for this format, and because Windows PE malware remains a major target of packer-based malware analysis. Therefore, our dataset is not intended to represent all malware ecosystems, such as ELF, APK, or script-based malware. Instead, it provides a large-scale, randomly sampled, and temporally diverse corpus for evaluating semantic correctness in Windows packer identification.

We integrated 81 unpackers into our framework, including both customized and generic unpackers. The customized unpackers were collected through online search and public repositories, including PyInstxtractor~\cite{pyinstxtractor}, PKLITE~\cite{pklite_data_unpacker}, UPX~\cite{upx}, ConfuserEx-Unpacker~\cite{confuserex_unpacker}, and Themida~\cite{themida_net_unpacker}. The generic unpackers include mal\_unpack~\cite{mal_unpack}, unipacker~\cite{unipacker}, and unpack~\cite{orcastor_unpack}. We also collected additional unpackers from Kanxue~\cite{kanxue_tools} and Awesome Executable Packing~\cite{awesome}, including Quick-Unpack~\cite{quick_unpack} and fuu~\cite{fuu_unpack}. For oracle construction, we combine complementary unpackers, including those available only as executable binaries, to obtain executable evidence of the underlying packing mechanisms. For repair construction, we inspect unpackers with available source code and extract detection predicates that can be evaluated directly on the input executable.

\subsection{RQ1: Correctness of Packer Identification}
\label{subsec:family-correct}
When evaluating packer identification tools, we prioritize recall  because these tools primarily serve as front-end filters for downstream malware analysis tasks, such as malware classification. Missing a packed sample means that the binary may bypass unpacking and enter subsequent analysis stages in its still-obfuscated form, which can directly degrade downstream results. Therefore, our main objective is to maximize the coverage of packed-binary recognition.

\subsubsection{Packer Family Identification}

\begin{table}[!t]
\centering
\caption{Packer identification recall (\%) for VirusTotal tools.}
\label{tab:closedsource}
\scriptsize
\setlength{\tabcolsep}{2pt}
\renewcommand{\arraystretch}{1.08}
\begin{tabular*}{\columnwidth}{@{\extracolsep{\fill}}lrlr@{}}
\toprule
Tool & Recall & Tool & Recall \\
\midrule
VT PEiD    & 46.8 & VT F-PROT  & 0.8  \\
VT Cyren   & 18.7 & VT Command & 0.1  \\
VT Varist  &  7.9 & VT Taggant & 0.02 \\
\bottomrule
\end{tabular*}
\vspace{-2mm}
\end{table}

\begin{table}[!t]
\centering
\caption{Packer identification recall (\%) for publicly available tools.}
\label{tab:opensource}
\scriptsize
\setlength{\tabcolsep}{2pt}
\renewcommand{\arraystretch}{1.08}
\begin{tabular*}{\columnwidth}{@{\extracolsep{\fill}}lrlr@{}}
\toprule
Tool (Strategy) & Recall & Tool (Strategy) & Recall \\
\midrule
DIE (hybrid)               & 48.0 & ReadPE (signature)     & 35.7 \\
Manalyze (signature)       & 48.9 & qu1cksc0pe (signature) & 49.6 \\
PEiD (signature)           & 42.9 & TrID (signature)       & 29.4 \\
PackHero (graph and learning)  & 42.6  &PyPEiD (signature)         & 38.5   \\
PyPackerDetect (signature) & 45.7 &  &  \\
\bottomrule
\end{tabular*}
\vspace{-5mm}
\end{table}

Table~\ref{tab:closedsource} and Table~\ref{tab:opensource} show that existing packer identification tools exhibit substantial limitations in packer family identification. VirusTotal tools generally achieve low recall, with VT PEiD performing best at 46.8\%, followed by VT Cyren at 18.7\% and VT Varist at 7.9\%. The remaining three tools, VT F-PROT, VT Command, and VT Taggant, achieve recall below 1\%, indicating negligible identification coverage in our evaluation. In contrast, the non-VirusTotal tools generally provide broader coverage. Qu1cksc0pe achieves the highest recall at 49.6\%, followed by Manalyze at 48.9\%, DIE at 48.0\%, and PyPackerDetect at 45.7\%. Nevertheless, even the best-performing tool correctly identifies fewer than half of the packed samples, highlighting the substantial limitations of existing packer family identification methods.

Beyond these aggregate results, we examine family-level identification performance separately for PE32 and PE64 binaries to characterize differences in packer coverage and identification effectiveness across executable architectures. For PE32 binaries, VirusTotal tools collectively recognize only 6 of the 14 evaluated major packer families. The remaining eight families, namely Themida~\cite{themida_oreans}, WinUpack~\cite{aldeid_winupack}, Petite~\cite{petite}, Armadillo~\cite{ugarte2016rambo}, MPRESS~\cite{mpress}, PESpin~\cite{securityxploded_pespin_plugin}, PyInstaller~\cite{pyinstaller_manual}, and Confuser~\cite{confuserex_github}, are not identified by any VirusTotal tool. Among the supported families, UPX~\cite{upx} receives the broadest support, being recognized by five of the six tools, followed by PECompact~\cite{PECompact} with four. ASPack~\cite{aspack_downloads}, MEW~\cite{mew_softpedia}, and NSPack~\cite{nsis} are each recognized by three tools, whereas Molebox~\cite{molebox_website} is supported by only two. Furthermore, none of the evaluated VirusTotal tools successfully identify packer families for PE64 binaries. However, support breadth does not necessarily translate into strong identification performance. For UPX, VT PEiD achieves 65.0\% recall, compared with 30.3\% for VT Cyren and 14.0\% for VT Varist. For ASPack, VT PEiD achieves 48.6\% recall, while the remaining tools achieve at most 0.5\%.  The highest recall for Molebox is only 28.6\%, achieved by VT PEiD, whereas PECompact reaches at most 15.4\% with VT Cyren. Identification performance is even weaker for MEW and NSPack, with recall not exceeding 0.6\% and 0.2\%, respectively. 

The evaluated non-VirusTotal tools provide broader family coverage on PE32 binaries, although their effectiveness remains highly uneven. For PE32 binaries, 11 of the 14 evaluated major packer families are identified by at least one tool, whereas WinUpack, PESpin, and Confuser remain undetected. UPX and PECompact are each identified by all nine tools, ASPack and MEW by eight, Molebox by seven, Armadillo and NSPack by six, and MPRESS by four. In contrast, Themida and Petite are each identified by only two tools, while PyInstaller is identified exclusively by DIE. Even among widely supported families, recall varies substantially. For UPX, qu1cksc0pe achieves 77.4\% recall, followed by Manalyze at 77.0\%, PackHero's majority-voting strategy at 74.5\%, and PyPackerDetect at 73.7\%. For ASPack, Manalyze achieves the highest recall at 85.4\%, followed by DIE at 65.9\%, while the remaining tools achieve at most 47.3\%. For MEW, qu1cksc0pe achieves 100\% recall, followed by PyPackerDetect and Manalyze at 82.9\% and 82.8\%, respectively, whereas the other five tools achieve recall of approximately 0.6\% or less. DIE performs particularly well on PyInstaller, Molebox, and MPRESS, achieving 97.1\%, 100\%, and 96.9\% recall, respectively. PEiD, Manalyze, PyPackerDetect, and qu1cksc0pe also achieve 85.7\% recall for Molebox. However, performance deteriorates considerably for several other families. The highest recall is only 36.6\% for Petite, below 25\% for PECompact, and approximately 14.8\% for Armadillo. Themida and NSPack remain particularly challenging, with the highest recall reaching only approximately 3.0\% and 0.2\%, respectively. These findings indicate that the broader coverage of these tools is driven by strong identification capabilities for a subset of packer families rather than consistently reliable performance across all families.

For PE64 binaries, packer family identification exhibits different coverage patterns and remains highly dependent on the tool and packer family. None of the evaluated VirusTotal tools correctly identifies any packer family. Among the nine evaluated non-VirusTotal tools, only 6 of the 10 evaluated major families are identified by at least one tool, whereas NSPack, PESpin, Petite, and WinUpack remain undetected. UPX is identified by four tools, Armadillo and Themida by three each, and MPRESS, PECompact, and PyInstaller by only one tool each. Despite this limited coverage, several tools achieve high recall for specific families. DIE achieves 100\% recall for UPX and MPRESS and 98.1\% for PyInstaller. It also performs best on Themida, with 48.7\% recall, followed by PEiD at 30.2\%. For UPX, TrID achieves 98.3\% recall, PEiD reaches 91.9\%, and PackHero's majority-voting strategy achieves 90.1\%. PackHero additionally identifies 55.6\% of PECompact samples using its clustering and mean strategies, whereas the other evaluated tools fail to identify this family. However, Armadillo remains poorly detected, with the highest recall reaching only 9.6\% for PEiD, PyPEiD, and qu1cksc0pe.

\subsubsection{Packedness Identification}
\begin{table}[!t]
\centering
\caption{Performance of heuristic-based strategies.}
\label{tab:combined_packed_detection_all_strategies}
\scriptsize
\setlength{\tabcolsep}{2pt}
\renewcommand{\arraystretch}{1.08}
\begin{tabular*}{\columnwidth}{@{\extracolsep{\fill}}lrlr@{}}
\toprule
Tool (Strategy) & Recall & Tool (Strategy) & Recall \\
\midrule
Bintropy (m0)     & 21.7 & REMINDer (heur1)       & 46.3 \\
Bintropy (m1)     & 14.2 & ReadPE (heur1)         & 39.4 \\
Bintropy (m0/m1)  & 25.8 & ReadPE (heur2)         & 86.8 \\
Bintropy (m0\&m1) & 10.1 & qu1cksc0pe (heur1)     & 99.1 \\
\midrule
PyPEiD (heur1)    & 46.5 & PyPackerDetect (heur1) & 82.3   \\
 Manalyze (heur1)     & 83.3  &                  &      \\
\bottomrule
\end{tabular*}
\vspace{-5mm}
\end{table}

Table~\ref{tab:combined_packed_detection_all_strategies} shows that even packedness identification, a coarser task than packer family identification, remains challenging for heuristic-based tools. Consistent with our recall-oriented evaluation objective, we focus on their ability to recognize packed binaries and minimize missed detections. Qu1cksc0pe (heur1) achieves the highest recall at 99.1\%, providing near-complete coverage. ReadPE (heur2) follows with 86.8\% recall, while Manalyze (heur1) and PyPackerDetect (heur1) achieve 83.3\% and 82.3\%, respectively. The remaining strategies exhibit substantially lower recall. PyPEiD (heur1), REMINDer (heur1), and ReadPE (heur1) detect 46.5\%, 46.3\%, and 39.4\% of packed samples, respectively.  Bintropy variants perform worst, with recall ranging from 10.1\% to 25.8\%. These results indicate that several existing heuristics miss a substantial proportion of packed executables when used independently.

Identification performance also varies across executable architectures. On PE32 binaries, Qu1cksc0pe achieves the highest recall at 99.1\%, followed by ReadPE (heur2) at 86.9\%, PyPackerDetect (heur1) at 83.5\%, and Manalyze (heur1) at 83.1\%. On PE64 binaries, Qu1cksc0pe remains highly effective, detecting 98.7\% of packed samples. Manalyze (heur1) and ReadPE (heur2) also maintain relatively high recall at 86.9\% and 85.5\%, respectively. In contrast, PyPackerDetect drops substantially from 83.5\% on PE32 to 51.6\% on PE64, while REMINDer (heur1) decreases from 47.1\% to 25.4\%. These differences indicate that heuristic effectiveness depends not only on the detection strategy but also on executable architecture. Nevertheless, the impact of architecture is not uniform across tools or packer families.

\begin{table}[!t]
\centering
\caption{Performance of rules in packer identification tools.}
\label{tab:overall_packed_detection_rule_based_manalyze}
\scriptsize
\setlength{\tabcolsep}{2pt}
\renewcommand{\arraystretch}{1.0}
\resizebox{\columnwidth}{!}{%
\begin{tabular}{@{}lr@{\hspace{8pt}}lr@{}}
\toprule
Strategy & Recall & Strategy & Recall \\
\midrule
\multicolumn{2}{l}{\textbf{Manalyze}} &
\multicolumn{2}{l}{\textbf{qu1cksc0pe}} \\
high\_entropy    & 68.6 & HasModified\_DOS\_Message & 27.5 \\
low\_imports     & 35.3 & HasOverlay                & 87.7 \\
resources\_size  & 0.0  & ImportTableIsBad          & 0.5 \\
rich\_header     & 5.1  & IsBeyondImageSize         & 0.5 \\
section\_name    & 77.0 & IsPacked                  & 46.5 \\
wx\_section      & 62.3 & string\_name\_match       & 74.6 \\
\midrule
\multicolumn{2}{l}{\textbf{PyPackerDetect}} &
\multicolumn{2}{l}{\textbf{ReadPE}} \\
bad\_ep\_sections     & 78.9 & high\_entropy  & 46.5 \\
low\_imports          & 42.4 & section\_name  & 79.7 \\
packer\_section\_match & 62.6 &                &      \\
unknown\_sections  & 66.2 &                &      \\
\bottomrule
\end{tabular}%
}
\vspace{-5mm}
\end{table}

We next examine the family-level performance of entropy-based heuristics, including Bintropy, REMINDer, and PyPEiD. Among these methods, PyPEiD (heur1), which applies a fixed entropy threshold of 7.0, provides relatively broad family-level coverage. It achieves substantial recall for Themida (73.3\%), Armadillo (62.0\%), PECompact (61.7\%), NSPack (62.8\%), PESpin (65.4\%), and Confuser (50.0\%), while reaching 100\% recall for PyInstaller and Molebox. REMINDer (heur1) performs better on UPX, achieving 65.4\% recall compared with 47.9\% for PyPEiD, but remains much weaker on ASPack (0.8\%), MEW (0.5\%), and PyInstaller (7.3\%). Bintropy generally exhibits lower recall, although its original combined heuristic detects 42.9\% of Themida, 35.4\% of Armadillo, 45.1\% of PECompact, 43.5\% of NSPack, and 44.6\% of PESpin samples. It also performs strongly on PyInstaller (99.1\%) and Molebox (85.7\%). These results demonstrate that entropy-based detection is highly dependent on packer-specific characteristics and thresholding strategies. For example, ASPack remains difficult for all three methods, with recall ranging from 0.8\% to 9.0\%. MEW is similarly challenging, with PyPEiD achieving only 1.4\% recall. In contrast, PyInstaller is nearly perfectly detected by Bintropy and PyPEiD but poorly detected by REMINDer. Thus, entropy alone does not provide a consistently reliable indicator of packedness across different packer families.

We further decompose the multi-rule heuristics in Manalyze, PyPackerDetect, Qu1cksc0pe, and ReadPE to identify the contributions of individual detection rules. As shown in Table \ref{tab:overall_packed_detection_rule_based_manalyze}, 
rule-level effectiveness varies substantially. In Manalyze, section\_name and high\_entropy achieve the highest recall at 77.0\% and 68.6\%, respectively, whereas resources\_size detects no packed samples and rich\_header achieves only 5.1\% recall. In PyPackerDetect, bad\_ep\_sections achieves 78.9\% recall, followed by unknown\_sections at 66.2\% and packer\_section\_match at 62.6\%, while low\_imports achieves 42.4\%. In Qu1cksc0pe, HasOverlay and string\_name\_match provide the broadest coverage, achieving 87.7\% and 74.6\% recall, respectively. In contrast, ImportTableIsBad and IsBeyondImageSize each achieve only 0.5\% recall. ReadPE exhibits a similar imbalance, with section\_name achieving 79.7\% recall compared with 46.5\% for high\_entropy. These findings indicate that the coverage of multi-rule heuristics is largely driven by a subset of strong detection rules, while several other rules contribute limited additional coverage.

Family-level analysis further reveals that these rules capture complementary packing artifacts. For ReadPE, section\_name achieves nearly perfect recall for WinUpack (100\%), MPRESS (99.4\%), and MEW (100\%), and remains effective for UPX (89.9\%). However, its recall drops to 33.4\% for Armadillo and 8.0\% for Confuser. In contrast, high\_entropy performs better on Themida (73.3\%), Armadillo (62.0\%), PECompact (61.7\%), NSPack (62.7\%), and PESpin (65.4\%), but is less effective for UPX (47.9\%) and MEW (1.4\%). These differences demonstrate that even within the same tool, distinct rules detect different subsets of packed executables.

A similar pattern appears in Qu1cksc0pe. HasOverlay achieves high recall for UPX (91.5\%), ASPack (95.8\%), PECompact (92.3\%), MEW (99.4\%), NSPack (90.0\%), and PESpin (90.7\%). In contrast, string\_name\_match is particularly effective for WinUpack (100\%), ASPack (99.9\%), MPRESS (99.8\%), and Confuser (100\%), but performs poorly on PyInstaller (7.0\%) and Molebox (14.3\%). Although the two rules exhibit overlapping coverage, their family-dependent effectiveness indicates that they capture different forms of packing evidence.

PyPackerDetect also exhibits complementary rule behavior. Its bad\_ep\_sections rule provides broad coverage across UPX (85.7\%), Themida (77.0\%), WinUpack (100\%), MPRESS (99.5\%), MEW (100\%), NSPack (91.5\%), and PESpin (93.9\%). The packer\_section\_match rule is similarly effective for WinUpack (100\%), MPRESS (99.2\%), and MEW (100\%), but achieves only 6.8\% recall for PyInstaller and fails to detect Confuser. In Manalyze, section\_name is highly effective for WinUpack (100\%), Themida (84.0\%), Petite (84.5\%), NSPack (95.0\%), PESpin (96.7\%), MEW (100\%), and MPRESS (99.4\%). Meanwhile, high\_entropy performs better for Armadillo (78.1\%) and PECompact (80.8\%), while maintaining high recall for MEW (99.8\%) and NSPack (92.3\%). These differences explain why combining complementary rules can achieve broader detection coverage than relying on individual rules.

\subsection{RQ2: Effectiveness of Fix Strategies}

\subsubsection{Packer Family Identification} 


The preceding evaluation identifies substantial family-dependent false negatives in signature-based packer identification tools. We investigate whether these failures can be recovered through two cumulative repair stages: (i) integrating complementary signatures already present in the evaluated tools and (ii) supplementing these signatures with static identification predicates extracted from publicly available unpacker implementations. We construct the rules on the initially analyzed dataset and evaluate the repair configuration on a separate test dataset.

\paragraph{Failure mechanisms and repair design}
The results in section \ref{subsec:family-correct} reveal two recurring limitations in existing packer identification tools. First, family coverage is fragmented: different tools recognize different subsets of packer families, and several families remain undetected by any evaluated tool. Second, identification effectiveness varies substantially across packer families. Even when a tool identifies some samples from a family, its recall may remain low. These findings motivate the subsequent investigation of complementary signatures and unpacker-derived detection logic to recover identification evidence missing from existing tools.

\paragraph{Two-stage repair mechanism}
Let $t$ denote an evaluated tool, $f$ a packer family, and $x$ an executable. Let $H_{t,f}^{(0)}(x)\in\{0,1\}$ indicate whether the unmodified tool predicts family $f$ for executable $x$. Let $\mathcal{R}^{\mathrm{sig}}_f$ be the collected family-$f$ signatures, $\mathcal{R}^{\mathrm{unp}}_f$ the family-$f$ static predicates extracted from unpacker implementations. Each rule $r$ or $s$ returns a Boolean match. We define the shared repaired decisions as
\begin{align}
H_{t,f}^{(1)}(x) &= \bigvee_{r\in\mathcal{R}^{sig}_{f}}r(x),
\label{eq:sig_union_revised}\\
H_{t,f}^{(2)}(x) &= H_{t,f}^{(1)}(x)\lor\bigvee_{r\in\mathcal{R}^{\mathrm{unp}}_f}r(x).
\label{eq:unpack_union_revised}
\end{align}
Equivalently, the first stage supplements missing signatures  by the validated family-specific signature pool $\mathcal{R}^{\mathrm{sig}}_f$ on each sample, whereas the second stage introduces structural evidence beyond the existing signature pool. These are common pooled-rule configurations, not individually validated patches applied to seven separate tool codebases. Accordingly, stage-1 and stage-2 results are identical across the tools by construction.

The first stage combines validated, family-labeled signature matches from the seven evaluated tools. On the initially analyzed dataset, this pool attains 63.2\% overall recall. Family recall reaches 86.3\% for UPX, 92.4\% for ASPack, and 100\% for MEW, demonstrating the value of complementary coverage. However, the union still detects none of the WinUpack, PESpin, or Confuser positives, and identifies only 8.6\% of Themida, 35.5\% of Petite, and 0.2\% of NSPack positives. Consequently, signature pooling alone cannot recover families for which discriminative evidence is missing from the source signatures. This result isolates the portion of missed detections attributable to fragmented signature coverage.

We next examine packer-identification conditions in public unpacker implementations to identify additional detection criteria, which we formalize as predicates. These predicates inspect PE sections, imports, raw-file markers, mapped entry-point (EP) windows, and structured CLR metadata. The extracted predicates operate directly on the input file. Their evaluation does not invoke the unpacker or execute the sample. Their role is to recover evidence not represented by the signature pool.
Themida illustrates the incremental benefit. Its signature union identifies only 8.6\% positives. Predicates derived from Unlicense and Unpack check Themida-related section names, imports,  section-prefix, file markers, and EP-stub patterns. With these predicates, 97.1\% of Themida positives are detected. Similarly, NSPack improves from 0.2\% positives after signature pooling to 94.6\% after incorporating section, marker, and stub checks and counting relative-call, jump, and conditional-branch bytes from Unpack. For Confuser, the repair instead derives predicates from ConfuserEx-Unpacker to locate structured CLR evidence, including a \texttt{koi} entry in the File table, \texttt{ConfusedByAttribute}, scoped version strings, selected assembly names, and an attribute combination associated with Confuser/DotNetPatcher. This identifies 88\% of Confuser samples, which were entirely missed by the signature union.

Table~\ref{tab:family_repair_summary} compares the seven original tools with the two cumulative shared repair configurations. On the initially analyzed dataset, the signature union achieves 63.2\% overall recall; adding unpacker-derived predicates increases this to 90.9\%. Applying the configuration to our test dataset yields 65.2\% and 89.6\% recall after stage-1 and stage-2, respectively. Thus, the second stage recovers substantial additional family evidence beyond the signature pool, and this gain persists on unseen samples.

\paragraph{Post Repair Residual Analysis}
For each family $f$, let $D_f$ be the set of ground-truth-positive executables. Because the repaired decision is shared, the residual false negatives are

\begin{equation}
\mathcal{M}^{(2)}_f=\{x\in D_f:H_f^{(2)}(x)=0\}.
\label{eq:signature_residual_revised}
\end{equation}

\begin{table}[!t]
\centering
\caption{Original, stage-1, and stage-2 packer family recall.}
\label{tab:family_repair_summary}
\scriptsize
\setlength{\tabcolsep}{2pt}
\renewcommand{\arraystretch}{1.08}
\begin{tabular*}{\columnwidth}{@{\extracolsep{\fill}}lrrrrrr@{}}
\toprule
& \multicolumn{3}{c}{Initially analyzed dataset}
& \multicolumn{3}{c@{}}{Test dataset} \\
\cmidrule(lr){2-4}
\cmidrule(l){5-7}
Tool & Original & Stage 1 & Stage 2
     & Original & Stage 1 & Stage 2 \\
\midrule
DIE        & 48.0 & 63.2 & 90.9  & 40.4 & 65.2 & 89.6\\
Manalyze       & 48.9  & 63.2 & 90.9 & 49.7 & 65.2 & 89.6 \\
PEiD         & 42.9  & 63.2 & 90.9 & 39.8 & 65.2 & 89.6 \\
PyPEiD       & 38.5  & 63.2 & 90.9 & 37.5 & 65.2 & 89.6 \\
PyPackerDetect & 45.7 & 63.2 & 90.9 & 50.4 & 65.2 & 89.6 \\
ReadPE     & 35.7 & 63.2 & 90.9 & 31.2 & 65.2 & 89.6 \\
Qu1cksc0pe & 49.6  & 63.2 & 90.9 & 64.3 & 65.2 & 89.6 \\
\bottomrule
\end{tabular*}
\vspace{-5mm}
\end{table}

We examine the residual samples to characterize the static evidence available after both repair stages. The following examples illustrate two recurring patterns observed in this analysis.

First, residual samples often lack consistent family-specific byte patterns. For example, the examined residual PESpin samples exhibit different entry-point (EP) layouts and share neither recognizable section names nor common EP signatures. In the analyzed residual Petite and UPX samples, the longest exact sequences shared across their respective EP windows contain only four and two zero bytes, respectively. These short padding sequences provide little discriminative value.

Second, shared patterns often reflect generic runtime structures rather than packer-specific code. For example, the EP prefix \texttt{FF 25 00 20 40 00} in the examined residual Confuser samples corresponds to a standard CLR startup jump. Similarly, a 17-byte sequence (\texttt{8B 44 24 10 89 6C 24 10 8D 6C 24 10 2B E0 53 56 57}) shared by a subset of examined residual Petite samples represents stack-frame and exception-handling initialization. Other Petite examples begin with \texttt{E8 29 04 00 00 E9 A9 FD FF FF}, which transfers control to compiler-runtime initialization. These patterns do not reliably distinguish packer families.

These observations help explain the remaining gap in static identification coverage. Dynamic unpacking can complement file-level checks with evidence exposed during execution. For example, a customized ASPack unpacker can proceed after an unmatched preliminary static check and identify packer structures in reconstructed process memory. Accessing such evidence entails sample execution, runtime inspection, and content recovery; identifying an appropriate unpacker may also require repeated attempts. Our framework uses unpacking outcomes to expose semantic errors and guide repairs that expand coverage within the static analysis setting. By improving identification before these recovery operations, the repaired tools narrow unpacker selection and reduce the need for trial-and-error unpacking.

\subsubsection{Packedness Identification}
The preceding evaluation reveals substantial false negatives in heuristic-based packedness detection. We investigate whether these failures can be recovered through two cumulative repair stages: (i) integrating complementary heuristic rules already present in the evaluated tools and (ii) supplementing these rules with predicates extracted from publicly available unpacker implementations. We construct the repair configurations using the initially analyzed dataset and evaluate them on a separate test dataset.

\paragraph{Failure mechanisms and repair design}

Bintropy, REMINDer, and PyPEiD illustrate how entropy-based
decisions can miss packed samples. Bintropy accepts a sample
when either of its two modes, m0 (whole-file) and m1 (section),
satisfies both the average- and maximum-entropy thresholds.
However, many packed samples satisfy the maximum-entropy
condition without satisfying the average-entropy condition
(Table~\ref{tab:bintropy_before_entropy_threshold}).
For MEW under m0, for example, 99.5\% of samples meet the
maximum-entropy threshold of 7.199, whereas only 0.2\% meet
the average-entropy threshold of 6.677. The alternative mode
does not adequately compensate for this gap. REMINDer similarly requires two conditions to hold
simultaneously: a writable entry-point (EP) section and
EP-section entropy above 6.85. As
Table~\ref{tab:reminder_original_rule_decomposition} shows,
MEW and WinUpack almost always satisfy the writability
condition but rarely satisfy the entropy condition, whereas
62.0\% of Confuser samples satisfy the entropy condition
but none satisfy the writability condition. Thus, the
conjunction excludes samples exhibiting only one of
these indicators. PyPEiD instead relies on a fixed entropy threshold, which can be affected by zero-byte regions and overlay content (Table~\ref{tab:pypeid_original_entropy_dilution_average_features}). Together, these cases show that a fixed entropy
criterion, whether applied globally or combined with
additional conditions, does not capture all observed
packing artifacts.

The remaining tools expose a related limitation:
combining multiple native rules still leaves detection
gaps when the rules provide insufficient complementary
coverage. Manalyze relies heavily on its section-name and high-entropy rules, while several other rules contribute limited coverage. PyPackerDetect exhibits the same problem more directly (Table~\ref{tab:pypackerdetect_original_strategy_decomposition}): none of its four examined rules detects Confuser, and each covers less than 10\% of PyInstaller samples. ReadPE's native entropy and section-name rules also have limited coverage for families such as ASPack, for which they detect only 9.0\% and 17.1\% of samples, respectively. Even Qu1cksc0pe, despite its high overall recall, misses samples in some families, including NSPack and PESpin. These observations indicate that the main limitation
lies in the evidence captured by individual heuristic
rules and their combinations. Nevertheless, different
tools exhibit complementary detection capabilities
across packer families. This complementarity motivates
the first repair stage, while the residual detection
gaps motivate the second stage based on unpacker-derived
rules.

\begin{table}[!t]
\centering
\caption{Bintropy before-fix entropy-threshold distribution by family
under modes m0 (whole-file) and m1 (section).}
\label{tab:bintropy_before_entropy_threshold}
\scriptsize
\setlength{\tabcolsep}{2pt}
\renewcommand{\arraystretch}{1.08}
\begin{tabular*}{\columnwidth}{@{\extracolsep{\fill}}lrrrrrr@{}}
\toprule
& \multicolumn{2}{c}{Avg $\geq 6.677$}
& \multicolumn{2}{c}{Max $\geq 7.199$}
& \multicolumn{2}{c@{}}{Both} \\
\cmidrule(lr){2-3}
\cmidrule(lr){4-5}
\cmidrule(l){6-7}
Family & m0 & m1 & m0 & m1 & m0 & m1 \\
\midrule
ASPack    &  5.3 &  6.6 & 97.2 &  8.4 &  5.3 &  5.5 \\
Armadillo & 13.9 & 37.4 & 79.5 & 49.5 & 13.9 & 34.1 \\
MEW       &  0.2 &  0.0 & 99.5 &  0.0 &  0.2 &  0.0 \\
MPRESS    & 31.2 & 32.4 & 86.1 & 65.9 & 31.2 & 26.6 \\
WinUpack  &  4.1 &  2.2 & 73.8 & 17.4 &  3.8 &  1.2 \\
\bottomrule
\end{tabular*}
\vspace{-3mm}
\end{table}

\begin{table}[!t]
\centering
\caption{Decomposition of REMINDer's original heuristic for
low-recall packer families.}
\label{tab:reminder_original_rule_decomposition}
\scriptsize
\setlength{\tabcolsep}{2pt}
\renewcommand{\arraystretch}{1.08}
\begin{tabular*}{\columnwidth}{@{\extracolsep{\fill}}lrrr@{}}
\toprule
Family & EP Writable & Entropy $> 6.85$ & EP Writable $\wedge$ Entropy \\
\midrule
ASPack      &  68.6 &  1.2 &  0.8 \\
Confuser  &   0.0 & 62.0 &  0.0 \\
MEW         & 100.0 &  0.5 &  0.5 \\
MPRESS      &  99.4 & 24.0 & 24.0 \\
PyInstaller &   7.3 &  7.7 &  7.3 \\
WinUpack    &  99.7 & 12.6 & 12.5 \\
\bottomrule
\end{tabular*}
\vspace{-3mm}
\end{table}

\begin{table}[!t]
\centering
\caption{Entropy dilution for low-recall PyPEiD families.
Ratios are computed relative to each sample's whole file size
and averaged within each family. }
\label{tab:pypeid_original_entropy_dilution_average_features}
\scriptsize
\setlength{\tabcolsep}{2pt}
\renewcommand{\arraystretch}{1.08}
\begin{tabular*}{\columnwidth}{@{\extracolsep{\fill}}lrr@{}}
\toprule
Family & Zero-byte Ratio & Overlay Ratio  \\
\midrule
MEW       & 28.1 & 78.9  \\
WinUpack  & 30.8 & 33.9 \\
MPRESS    & 40.6 & 27.0 \\
PESpin    & 19.2 & 45.6  \\
Petite    & 16.4 & 36.7  \\
Themida   & 12.8 & 58.1  \\
\bottomrule
\end{tabular*}
\vspace{-3mm}
\end{table}

\paragraph{Two-stage repair mechanism.}
Let $H_t^{(0)}(x)\in\{0,1\}$ denote tool $t$'s original
packedness decision for sample $x$, where 1 indicates
packed. Let $\mathcal{R}_t$ be the set of complementary
heuristic rules imported from other tools. The first-stage
decision is defined as
\begin{equation}
H_t^{(1)}(x)=H_t^{(0)}(x)\lor
\bigvee_{r\in\mathcal{R}_t}r(x).
\label{eq:packedness_stage1}
\end{equation}
The original heuristic remains a complete decision
component: its thresholds, constituent rules, and
internal combination logic are unchanged. Thus,
the first stage extends the original decision with
complementary evidence rather than modifying its
native detection conditions.

To guide the second stage, we analyze the remaining
false negatives by packer family to identify detection
evidence not adequately represented by the first-stage
heuristic pool. We then examine detection conditions
in publicly available unpacker implementations and
extract supplementary rules for poorly detected families. Let $\mathcal{F}_{\mathrm{res}}$ denote the families
targeted by residual analysis, and let
$\mathcal{S}_f^{\mathrm{unp}}$ denote the set of
unpacker-derived supplementary rules for family $f$.
The second-stage decision is
\begin{equation}
H_t^{(2)}(x)=H_t^{(1)}(x)\lor
\bigvee_{f\in\mathcal{F}_{\mathrm{res}}}
\bigvee_{s\in\mathcal{S}_f^{\mathrm{unp}}}s(x).
\label{eq:packedness_stage2}
\end{equation}

\begin{table}[t]
\centering
\caption{Representative weak cases of PyPackerDetect's native rules.}
\label{tab:pypackerdetect_original_strategy_decomposition}
\scriptsize
\setlength{\tabcolsep}{2.3pt}
\resizebox{\columnwidth}{!}{%
\begin{tabular}{lrrrr}
\toprule
Family
& \texttt{packer\_sec}
& \texttt{bad\_ep}
& \texttt{low\_imports}
& \texttt{unknown\_sec}
 \\
\midrule
Confuser
& 0.00 & 0.00 & 0.00 & 0.00 \\
PyInstaller
& 6.80& 7.94 & 7.26 & 0.23  \\
 Armadillo 
& 17.0 &58.3 & 48.5 & 22.3  \\
Petite
& 15.7 & 42.0 & 46.1 & 9.8  \\
PECompact
& 26.7 & 68.9 & 64.4 & 14.9  \\
\bottomrule
\end{tabular}%
}
\vspace{-5mm}
\end{table}

For example, let \(C(x)\) and \(A(x)\) denote two unpacker-derived supplementary rules for Confuser and Armadillo, respectively. When these two rules are considered, the repaired decision can be expressed as

\[
H_t^{(2)}(x)=H_t^{(0)}(x)\lor
\bigvee_{r\in\mathcal{R}_t}r(x)
\lor C(x)\lor A(x).
\]

Both stages preserve earlier positive decisions.
Consequently, recall cannot decrease on the
datasets.

The first stage draws on three complementary rules:
\texttt{Qu1cksc0pe.HasOverlay},
\texttt{ReadPE.section\_name}, and
\texttt{Manalyze.high\_entropy}. Their effectiveness
varies across packing artifacts and families. The overlay rule provides complementary coverage
for families such as MEW and PyInstaller, while the
section-name rule is effective for MPRESS and WinUpack. Manalyze's entropy rule contributes additional
detections for families such as ASPack and Confuser. Complementarity also occurs within a feature category:
ReadPE's section-name rule captures patterns missed
by Manalyze's native section-name rule. Based on these complementary strengths, we augment Bintropy, REMINDer, and PyPackerDetect with all three rules. PyPEiD and Manalyze receive the overlay and ReadPE section-name rules. ReadPE receives the overlay and Manalyze entropy rules; and Qu1cksc0pe receives the ReadPE section-name rule. Each tool retains its original heuristic, with the selected complementary
rules combined according to Eq.~(\ref{eq:packedness_stage1}).

Table~\ref{tab:packedness_repair_summary} compares
the original and repaired packedness detection
performance. On the initially analyzed dataset, the first-stage repair increases recall from 25.8\%
to 99.5\% for Bintropy, from 46.3\% to 99.4\%
for REMINDer, and from 46.5\% to 99.3\% for
PyPEiD. Tools with broader native rule sets also
benefit: Manalyze, PyPackerDetect, and ReadPE reach
99.8\%, 99.8\%, and 99.5\% recall, respectively.
Qu1cksc0pe, which already achieves 99.1\% recall,
improves to 99.6\%. On the separate test dataset, first-stage recall
ranges from 97.6\% to 98.8\% across the seven
tools. These results demonstrate the effectiveness
of complementary heuristic integration in recovering
missed packed samples. However, the high aggregate
recall conceals residual weaknesses for certain
packer families, motivating the second stage.

\begin{table}[!t]
\centering
\caption{Original, stage-1, and stage-2 packedness recall.}
\label{tab:packedness_repair_summary}
\scriptsize
\setlength{\tabcolsep}{2pt}
\renewcommand{\arraystretch}{1.08}
\begin{tabular*}{\columnwidth}{@{\extracolsep{\fill}}lrrrrrr@{}}
\toprule
& \multicolumn{3}{c}{Initially analyzed dataset}
& \multicolumn{3}{c@{}}{Test dataset} \\
\cmidrule(lr){2-4}
\cmidrule(l){5-7}
Tool & Original & Stage 1 & Stage 2
     & Original & Stage 1 & Stage 2 \\
\midrule
Bintropy       & 25.8 & 99.5 & 99.6 & 27.6 & 98.2 & 99.1 \\
REMINDer       & 46.3 & 99.4 & 99.6 & 41.9 & 98.1 & 99.1 \\
PyPEiD         & 46.5 & 99.3 & 99.5 & 44.4 & 97.6 & 99.0 \\
Manalyze       & 83.3 & 99.8 & 99.9 & 82.2 & 98.8 & 99.5 \\
PyPackerDetect & 82.3 & 99.8 & 99.9 & 71.5 & 98.5 & 99.3 \\
Qu1cksc0pe     & 99.1 & 99.6 & 99.8 & 97.2 & 98.0 & 99.1 \\
ReadPE         & 86.8 & 99.5 & 99.6 & 80.5 & 98.2 & 99.1 \\
\bottomrule
\end{tabular*}
\vspace{-4mm}
\end{table}

Despite the first-stage augmentation, some families,
particularly Confuser and Armadillo, remain poorly
detected. The reused entropy, overlay, and section-name
rules provide partial and overlapping coverage, but
leave gaps in identifying protection artifacts at
the managed-code level or entry point. For Confuser, we extract a supplementary detection
rule from ConfuserEx-Unpacker to inspect Common
Language Runtime (CLR) metadata. Its indicators
include \texttt{ConfusedByAttribute}, scoped
\texttt{Confuser v}/\texttt{ConfuserEx v} strings,
known Bed's-mod assembly names, and a
Confuser/DotNetPatcher attribute combination.
Incorporating this rule enables all seven tools
to achieve 100\% recall for Confuser. Similarly, to address the remaining Armadillo
false negatives, we extract an Armadillo-oriented
detection rule from Unpack. This rule checks for
\texttt{60 E8} exactly at the mapped entry point,
or for the joint occurrence of
\texttt{E8 00 00 00 00} and \texttt{EB 00}
within the next 512 bytes. Incorporating this
rule increases Armadillo recall to 99.7\% and
improves overall recall by an additional
0.72--1.32\%. With these unpacker-derived rules, the second-stage
repair achieves 99.5--99.9\% recall on the initially
analyzed dataset and 99.0--99.5\% recall on the
separate test dataset.

\paragraph{Post Repair Residual Analysis} We also examine the residual false negatives in packedness detection. Representative examples exhibit weak or inconsistent packing indicators and generic runtime artifacts that provide insufficient discriminative evidence for the examined static rules. Combining complementary heuristics and unpacker-derived static predicates provides broader packedness coverage directly from the binary file.

\subsection{RQ3: Downstream Impact on Malware Classification}
To assess the practical consequences of packer identification errors, we evaluate their downstream impact on malware classification. Table~\ref{tab:accuracy1} compares the performance of 14 representative malware classification methods before and after repairing packer identification results and applying the corresponding unpackers. All evaluated methods exhibit improvements in precision, recall, and F1-score, although the magnitude of these improvements varies considerably. These results indicate that packer identification errors can adversely affect downstream malware classification. When a packed binary is not correctly identified, the analysis pipeline may skip unpacking or select an inappropriate unpacker, potentially leaving the original program content inaccessible. Such unresolved obfuscation can obscure discriminative features used for malware classification. Therefore, accurate packer identification and successful unpacking can play an important role in improving downstream classification performance, particularly for methods sensitive to packing-induced transformations.

Across the 14 evaluated methods, average classification precision improves by 13.6 percentage points after the fixes. Although this average is influenced by the unusually large improvement observed for DNN~\cite{Saxe}, the remaining 13 methods still achieve an average precision gain of 7.4 percentage points. Importantly, precision improvements are observed across all 14 classification methods, including those that already achieve relatively strong performance before repair. For example, Malfiner~\cite{zhong2024enhancing} improves from 96.0\% to 99.6\%, while VisUnpac~\cite{zhong2025unveiling} increases from 96.6\% to 99.7\%. Similarly, DRBA~\cite{Zhihua} improves from 90.2\% to 99.3\%, and GIST~\cite{Nataraj} increases from 89.4\% to 98.8\%. The consistent improvements across diverse classifiers demonstrate the downstream value of repairing packer identification errors: more reliable packer predictions guide appropriate unpacking, yielding more informative program representations for malware classification. Besides, the average precision of scan engines in VirusTotal \cite{virustotal} also improves from 14.2\% to 25.3\%.

The largest performance improvement is observed for DNN~\cite{Saxe}, whose precision increases from 0.01\% before the fixes to 95.3\% afterward. This substantial improvement demonstrates the pronounced impact that packer identification errors and unsuccessful unpacking can have on downstream malware classification. By enabling the recovery of analyzable program content, the repaired identification pipeline provides more informative representations for malware classification. KNN~\cite{Ouahab}, by comparison, achieves 84.8\% precision before the fixes and improves to 95.0\% afterward. Its before-fix performance is also uneven across malware categories, with precision below 50\% for more than 35\% of them. Together, these observations suggest that packer identification errors can affect different classification methods to substantially different degrees, depending on the features and representations used for malware classification.

\begin{table}[t!]
\centering
\captionsetup{skip=2pt}
\caption{Malware Classification Before vs. After Fix (\%)}
\renewcommand{\arraystretch}{1.15}
\begin{tabular}{|l|ccc|ccc|}
\hline
\multirow{2}{*}{\textbf{Method}} & \multicolumn{3}{c|}{\textbf{ Before Fix}} & \multicolumn{3}{c|}{\textbf{ After Fix}} \\
\cline{2-7}
& Prec. & Recall & F1 & Prec. & Recall & F1  \\
\hline
DNN \cite{Saxe} & 0.01 & 1.00 & 0.02  & 95.3 & 94.2 & 94.5 \\
KNN \cite{Ouahab} & 84.8 & 82.0 & 81.9  & 95.0 & 92.8 & 92.7  \\
CNN+DT \cite{ahmed2024novel} & 21.0 & 22.5 & 18.4  & 23.5 & 23.7 & 21.5  \\
DRBA \cite{Zhihua} & 90.2 & 89.0 & 87.5 & 99.3 & 99.3 & 99.3  \\
dataenhance \cite{liu2024efficient} & 91.1 & 90.2 & 89.9 & 98.9 & 98.8 & 98.9 \\
VisMal \cite{zhong2022malware} & 94.3 & 93.0 & 92.1  & 99.7 & 99.7 & 99.7 \\
GIST \cite{Nataraj} & 89.4 & 89.1 & 89.2  & 98.8 & 98.7 & 98.7 \\
MDMC \cite{Baoguo} & 92.4 & 92.2 & 91.4& 99.4 & 99.4 & 99.4 \\
LBP \cite{liu2019new} & 90.8 & 90.4 & 90.4 & 98.9 & 98.9 & 98.9 \\
Zhao \emph{et al.} \cite{zhao2023image} & 83.6 & 77.0 & 78.0 & 93.8 & 87.4 & 89.0 \\
Chen et al. \cite{chen2020malware} & 89.5 & 89.9 & 88.4 & 97.8 & 97.7 & 97.6 \\
Adkins et al. \cite{adkins2013heuristic} & 87.9 & 87.3 & 86.1  & 98.9 & 99.0 & 98.7  \\
Malfiner \cite{zhong2024enhancing} & 96.0 & 95.6& 95.3  & 99.6 & 99.6 & 99.6  \\
VisUnpac \cite{zhong2025unveiling} & 96.6 & 96.1 & 95.9 & 99.7 & 99.7 & 99.7  \\
\hline
\end{tabular}
\label{tab:accuracy1}
\vspace{-2mm}
\end{table}

\begin{figure}[ht]
    \centering
    \begin{subfigure}{0.47\linewidth}
        \centering
        \includegraphics[width=0.9\linewidth, height=1.2in]{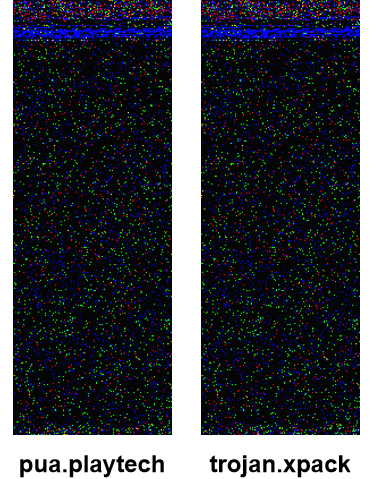}
        \caption{Packed Samples}
    \end{subfigure}
    \hfill
    \begin{subfigure}{0.47\linewidth}
        \centering
        \includegraphics[width=0.9\linewidth, height=1.2in]{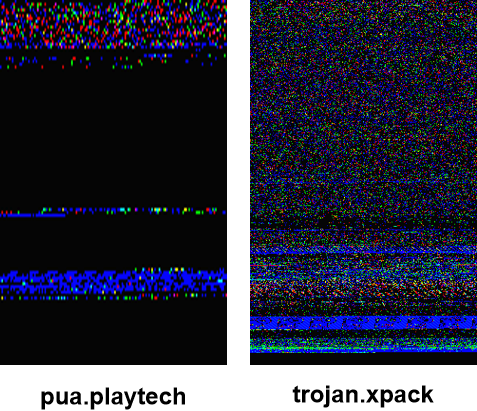}
        \caption{Unpacked Samples}
    \end{subfigure}
    \captionsetup{skip=3pt}
    \caption{VisUnpac Feature Data Distributions}
    \label{fig:feature}
    \vspace{-5mm}
\end{figure}

We further investigate this impact using VisUnpac~\cite{zhong2025unveiling}, one of the best-performing methods in Table~\ref{tab:accuracy1}. Although VisUnpac achieves relatively high performance before the fixes, its precision continues to increase from 96.6\% to 99.7\% after the fixes. To better understand the remaining errors, we examine the malware categories exhibiting the largest performance differences between the two settings. These categories include \textit{pua.playtech}, \textit{trojan.fareit}, \textit{trojan.cobaltstrike}, \textit{trojan.rozena}, \textit{trojan.toolbar}, \textit{trojan.swisyn}, \textit{trojan.wacapew}, and \textit{trojan.xpack}. Before the fixes, precision is as low as 36.8\% for \textit{pua.playtech}, 37.5\% for \textit{trojan.fareit}, 44.4\% for \textit{trojan.cobaltstrike}, 48.7\% for \textit{trojan.swisyn}, and 42.6\% for \textit{trojan.wacapew}, while \textit{trojan.rozena} is not correctly classified. After the fixes, all of these categories achieve 100\% precision. 

A closer examination of \textit{pua.playtech} and \textit{trojan.xpack} provides an illustrative example of this effect. Samples from both categories are packed with NSPack, and their packed feature representations exhibit similar distributions, as illustrated in Fig.~\ref{fig:feature}. This similarity is consistent with the possibility that shared packing artifacts obscure differences between the underlying malware categories, contributing to classification confusion. After the corresponding unpackers are applied, VisUnpac recovers more distinguishable feature distributions from the unpacked program content, consistent with the observed improvements in classification performance. This example illustrates how packing-related features can dominate the observable representations of different malware categories, while unpacking can help recover category-specific information.

Taken together, these results demonstrate that packer identification errors can propagate into downstream malware classification by preventing the recovery of discriminative program features. Correcting these errors improves classification performance across diverse methods and helps distinguish malware categories whose packed representations exhibit substantial similarity. These findings underscore the importance of semantically reliable packer identification for downstream malware analysis. More broadly, our findings challenge the common assumption in prior work \cite{Saxe,Ouahab,Nataraj,Yakura,zhong2022malware,liu2019new} that packing has only limited impact on malware classification.

\section{Threats to Validity}
\label{sec:validity}

\parhead{Internal Validity.} 
A primary threat is the correctness of our unpacking-validated oracle. Although successful unpacking provides strong semantic evidence, imperfect unpackers may still fail on correctly identified samples or only partially recover analyzable program content. We mitigate this threat by integrating an extensive collection of generic and customized unpackers into a unified pipeline and leveraging their successful unpacking outcomes to construct semantic test oracles. Nevertheless, the reliability of these oracles depends on the capabilities and coverage of the available unpackers. A further threat concerns downstream dataset labeling. Although labeling combined automatic assignment, database confirmation, dynamic analysis, reverse engineering, and multi-author agreement checks, some residual bias may remain. We mitigate this threat through iterative validation with Cohen's Kappa consistently exceeding 0.9 in subsequent annotation rounds and third-author adjudication.

\parhead{External Validity.} 
Our benchmark may not fully represent all real-world packers, malware distributions, or executable formats. We focus on Windows PE executables because the available packer identification tools and unpackers primarily target this format. In addition, our repairs focus on open-source signature-based and heuristic-based tools whose decision logic is accessible for analysis and modification. Consequently, the proposed repair strategies are not directly evaluated on proprietary tools or graph-based and learning-based methods requiring different modification approaches. Finally, unpacker-guided repair is limited by the availability of publicly accessible unpackers, although our framework is designed to incorporate additional unpackers in the future.

\section{Conclusion}
\label{sec:conclusion}
We present a semantic validation framework for packer identification tools that leverages unpackers as executable semantic contracts to expose and repair semantic errors. By constructing unpacking-validated test oracles and normalizing heterogeneous tool outputs, our framework enables systematic detection, localization, and repair of semantic faults without requiring manually labeled packer ground truth. Our empirical study reveals widespread identification failures caused by incomplete signatures and insufficient heuristic coverage. We further demonstrate that integrating complementary signatures and heuristic rules with static unpacker-derived detection predicates substantially improves packer identification coverage. The repaired rules operate directly on input executables, strengthening the front-end decisions that guide subsequent unpacking and malware analysis. Moreover, our downstream evaluation shows that correcting these errors improves malware classification performance across diverse classification methods. These findings underscore the importance of semantic correctness in foundational malware analysis tools and its role in ensuring reliable downstream malware analysis.

\bibliographystyle{IEEEtran}
\bibliography{sample-base}

@inproceedings{stephens2018probabilistic,
  title={Probabilistic obfuscation through covert channels},
  author={Stephens, Jon and Yadegari, Babak and Collberg, Christian and Debray, Saumya and Scheidegger, Carlos},
  booktitle={2018 IEEE European Symposium on Security and Privacy (EuroS\&P)},
  pages={243--257},
  year={2018},
  organization={IEEE}
}

@article{zhong2022malware,
  title={Malware-on-the-brain: Illuminating malware byte codes with images for malware classification},
  author={Zhong, Fangtian and Chen, Zekai and Xu, Minghui and Zhang, Guoming and Yu, Dongxiao and Cheng, Xiuzhen},
  journal={IEEE Transactions on computers},
  volume={72},
  number={2},
  pages={438--451},
  year={2022},
  publisher={IEEE}
}

@inproceedings{bhardwaj2021reverse,
  title={Reverse engineering-a method for analyzing malicious code behavior},
  author={Bhardwaj, Vivek and Kukreja, Vinay and Sharma, Chetan and Kansal, Isha and Popali, Renu},
  booktitle={2021 international conference on advances in computing, communication, and control (ICAC3)},
  pages={1--5},
  year={2021},
  organization={IEEE}
}

@inproceedings{li2025malmixer,
  title={Malmixer: Few-shot malware classification with retrieval-augmented semi-supervised learning},
  author={Li, Jiliang and Zhang, Yifan and Huang, Yu and Leach, Kevin},
  booktitle={2025 IEEE 10th European Symposium on Security and Privacy (EuroS\&P)},
  pages={268--288},
  year={2025},
  organization={IEEE}
}

@article{maleki2019improved,
  title={An improved method for packed malware detection using PE header and section table information},
  author={Maleki, Nahid and Bateni, Mehdi and Rastegari, Hamid},
  journal={International Journal of Computer Network and Information Security},
  volume={9},
  number={9},
  pages={9},
  year={2019},
  publisher={Modern Education and Computer Science Press}
}

@inproceedings{downing2021deepreflect,
  title={$\{$DeepReflect$\}$: Discovering malicious functionality through binary reconstruction},
  author={Downing, Evan and Mirsky, Yisroel and Park, Kyuhong and Lee, Wenke},
  booktitle={30th USENIX security symposium (USENIX Security 21)},
  pages={3469--3486},
  year={2021}
}

@misc{malscan_github,
  author       = {{Ice3man543}},
  title        = {{MalScan}},
  howpublished = {\url{https://github.com/Ice3man543/MalScan}},
  note         = {GitHub repository, accessed March 18, 2026}
}

@misc{peid_top4download,
  author       = {{Top4Download.com}},
  title        = {{PEiD screenshot}},
  howpublished = {\url{https://www.top4download.com/peid-tab/screenshot-gaqrbxek.html}},
  note         = {Accessed: 2026-03-18}
}

@article{zheng2025gupacker,
  title={Gupacker: Generalized Unpacking Framework for Android Malware},
  author={Zheng, Tao and Hou, Qiyu and Chen, Xingshu and Ren, Hao and Li, Meng and Li, Hongwei and Shen, Changxiang},
  journal={IEEE Transactions on Information Forensics and Security},
  year={2025},
  publisher={IEEE}
}

@misc{apppeid_github,
  author       = {{wolfram77web}},
  title        = {{app-peid}},
  howpublished = {\url{https://github.com/wolfram77web/app-peid}},
  note         = {GitHub repository. Accessed: 2026-03-14},
  year         = {2026}
}

@misc{asl_github,
  author       = {{Exeinfo ASL}},
  title        = {{ASL}},
  howpublished = {\url{https://github.com/ExeinfoASL/ASL/tree/master}},
  note         = {GitHub repository. Accessed: 2026-03-14},
  year         = {2026}
}

@misc{trid_website,
  author       = {Marco Pontello},
  title        = {{TrID - File Identifier}},
  howpublished = {\url{https://mark0.net/soft-trid-e.html}},
  note         = {Accessed: 2026-03-14}
}

@misc{qu1cksc0pe_github,
  author       = {{CYB3RMX}},
  title        = {{Qu1cksc0pe}},
  howpublished = {\url{https://github.com/CYB3RMX/Qu1cksc0pe}},
  note         = {GitHub repository. Accessed: 2026-03-14},
  year         = {2026}
}

@misc{pypackerdetect_github,
  author       = {{packing-box}},
  title        = {{pypackerdetect: Packing detection tool for PE files}},
  howpublished = {\url{https://github.com/packing-box/pypackerdetect?tab=readme-ov-file}},
  note         = {GitHub repository, accessed March 14, 2026}
}

@misc{ffri_pypeid,
  author       = {{FFRI Security, Inc.} and {Koh M. Nakagawa}},
  title        = {{pypeid: Yet another implementation of PEiD with yara-python}},
  howpublished = {\url{https://github.com/FFRI/pypeid}},
  note         = {GitHub repository, accessed March 14, 2026}
}

@misc{manalyze_github,
  author       = {Alexandre Borges},
  title        = {{Manalyze: A static analyzer for PE executables}},
  howpublished = {\url{https://github.com/JusticeRage/Manalyze}},
  year         ={2026},
  note         = {GitHub repository, accessed March 14, 2026}
}

@INPROCEEDINGS{Saxe,
	author={Saxe, Joshua and Berlin, Konstantin},
	booktitle={2015 10th International Conference on Malicious and Unwanted Software (MALWARE)}, 
	title={Deep neural network based malware detection using two dimensional binary program features}, 
	year={2015},
	volume={},
	number={},
	pages={11-20},
	doi={10.1109/MALWARE.2015.7413680}}

@InProceedings{Ouahab,
	author="Ben Abdel Ouahab, Ikram
	and Bouhorma, Mohammed
	and Boudhir, Anouar Abdelhakim
	and El Aachak, Lotfi",
	editor="Ben Ahmed, Mohamed
	and Boudhir, Anouar Abdelhakim
	and Santos, Domingos
	and El Aroussi, Mohamed
	and Karas, {\.{I}}smail Rak{\i}p",
	title="Classification of Grayscale Malware Images Using the K-Nearest Neighbor Algorithm",
	booktitle="Innovations in Smart Cities Applications Edition 3",
	year="2020",
	publisher="Springer International Publishing",
	address="Cham",
	pages="1038--1050",
	isbn="978-3-030-37629-1"
}

@inproceedings{Yakura,
  title={Malware analysis of imaged binary samples by convolutional neural network with attention mechanism},
  author={Yakura, Hiromu and Shinozaki, Shinnosuke and Nishimura, Reon and Oyama, Yoshihiro and Sakuma, Jun},
  booktitle={Proceedings of the Eighth ACM Conference on Data and Application Security and Privacy},
  pages={127--134},
  year={2018}
}

@inproceedings{ahmed2024novel,
  title={A Novel Approach to Malware Detection using Machine Learning and Image Processing},
  author={Ahmed, Saadaldeen Rashid and Mohamed, Salah J and Aljanabi, Mohammed S and Algburi, Sameer and Majeed, Duaa A and Kurdi, Neesrin Ali and Al-Sarem, Mohammed and Tawfeq, Jamal Fadhil},
  booktitle={Proceedings of the Cognitive Models and Artificial Intelligence Conference},
  pages={298--302},
  year={2024}
}

@ARTICLE{Zhihua,
	author={Cui, Zhihua and Xue, Fei and Cai, Xingjuan and Cao, Yang and Wang, Gai-ge and Chen, Jinjun},
	journal={IEEE Transactions on Industrial Informatics}, 
	title={Detection of Malicious Code Variants Based on Deep Learning}, 
	year={2018},
	volume={14},
	number={7},
	pages={3187-3196},
	doi={10.1109/TII.2018.2822680}}

@article{liu2024efficient,
  title={Efficient and generalized image-based CNN algorithm for multi-class malware detection},
  author={Liu, Yajun and Fan, Hong and Zhao, Jianguang and Zhang, Jianfang and Yin, Xinxin},
  journal={IEEE Access},
  year={2024},
  publisher={IEEE}
}

@inproceedings{Nataraj,
	author = {Nataraj, L. and Karthikeyan, S. and Jacob, G. and Manjunath, B. S.},
	title = {Malware images: visualization and automatic classification},
	year = {2011},
	isbn = {9781450306799},
	publisher = {Association for Computing Machinery},
	address = {New York, NY, USA},
	url = {https://doi.org/10.1145/2016904.2016908},
	doi = {10.1145/2016904.2016908},
	booktitle = {Proceedings of the 8th International Symposium on Visualization for Cyber Security},
	articleno = {4},
	numpages = {7},
	series = {VizSec '11}
}

@article{Baoguo,
	title = {Byte-level malware classification based on markov images and deep learning},
	journal = {Computers \& Security},
	volume = {92},
	pages = {101740},
	year = {2020},
	issn = {0167-4048},
	doi = {https://doi.org/10.1016/j.cose.2020.101740},
	author = {Baoguo Yuan and Junfeng Wang and Dong Liu and Wen Guo and Peng Wu and Xuhua Bao},
}

@inproceedings{adkins2013heuristic,
  title={Heuristic malware detection via basic block comparison},
  author={Adkins, Francis and Jones, Luke and Carlisle, Martin and Upchurch, Jason},
  booktitle={2013 8th International Conference on Malicious and Unwanted Software:" The Americas"(MALWARE)},
  pages={11--18},
  year={2013},
  organization={IEEE}
}

@article{virustotal,
 author  = {Darroch, Gordon},
 date    = {2026-02-14},
 title   = {How it works},
 journal = {VirusTotal},
 url     = {https://docs.virustotal.com/docs/how-it-works},
 year ={2026},
 urldate = {2026-02-14}
}

@inproceedings{chen2020malware,
  title={A malware classification method based on basic block and CNN},
  author={Chen, Jinrong},
  booktitle={Neural Information Processing: 27th International Conference, ICONIP 2020, Bangkok, Thailand, November 18--22, 2020, Proceedings, Part IV 27},
  pages={275--283},
  year={2020},
  organization={Springer}
}

@article{zhao2023image,
  title={Image-Based Malware Classification Method with the AlexNet Convolutional Neural Network Model},
  author={Zhao, Zilin and Zhao, Dawei and Yang, Shumian and Xu, Lijuan},
  journal={Security and Communication Networks},
  volume={2023},
  number={1},
  pages={6390023},
  year={2023},
  publisher={Wiley Online Library}
}

@article{liu2019new,
  title={A new learning approach to malware classification using discriminative feature extraction},
  author={Liu, Ya-shu and Lai, Yu-Kun and Wang, Zhi-Hai and Yan, Han-Bing},
  journal={IEEE Access},
  volume={7},
  pages={13015--13023},
  year={2019},
  publisher={IEEE}
}

@inproceedings{di2025packhero,
  title={PackHero: A Scalable Graph-based Approach for Efficient Packer Identification},
  author={Di Gennaro, Marco and D’Onghia, Mario and Polino, Mario and Zanero, Stefano and Carminati, Michele},
  booktitle={International Conference on Detection of Intrusions and Malware, and Vulnerability Assessment},
  pages={253--274},
  year={2025},
  organization={Springer}
}

@article{lyda2007using,
  title={Using entropy analysis to find encrypted and packed malware},
  author={Lyda, Robert and Hamrock, James},
  journal={IEEE security \& privacy},
  volume={5},
  number={2},
  pages={40--45},
  year={2007},
  publisher={IEEE}
}

@article{biondi2019effective,
  title={Effective, efficient, and robust packing detection and classification},
  author={Biondi, Fabrizio and Enescu, Michael A and Given-Wilson, Thomas and Legay, Axel and Noureddine, Lamine and Verma, Vivek},
  journal={Computers \& Security},
  volume={85},
  pages={436--451},
  year={2019},
  publisher={Elsevier}
}

@misc{upx,
    author ={ Markus F.X.J. Oberhumer and László Molnár and John F. Reiser.},
	title = {the Ultimate Packer for eXecutables
},
	howpublished = {\url{https://upx.github.io/}},
	note = {Accessed: 2024-12-30}
}

@misc{themida_oreans,
  author       = {{Oreans Technologies}},
  title        = {{Themida Protector}},
  howpublished = {\url{https://www.oreans.com/Themida.php}},
  note         = {Accessed: 2026-03-22}
}

@misc{aspack_downloads,
  author       = {{ASPack Software}},
  title        = {{ASPack Downloads}},
  howpublished = {\url{http://www.aspack.com/downloads.html}},
  year         ={2020},
  note         = {Accessed: 2026-03-22}
}

@misc{mew_softpedia,
  author       = {{Softpedia}},
  title        = {{MEW - Download - Softpedia}},
  howpublished = {\url{https://www.softpedia.com/get/Programming/Packers-Crypters-Protectors/MEW-SE.shtml}},
  year         ={2004},
  note         = {Accessed: 2026-03-22}
}

@misc{PECompact,
    author ={Jeremy Collake},
	title = {PECompact – Windows (PE) Executable Compressor
},
	howpublished = {\url{https://bitsum.com/portfolio/pecompact/}},
    year = {2017},
	note = {Accessed: 2026-03-26}
}

@misc{petite,
    author ={Un4seen Developments},
	title = {Win32 Executable Compressor
},
	howpublished = {\url{https://www.un4seen.com/petite/

}},
	note = {Accessed: 2026-03-26}
}

@inproceedings{ugarte2016rambo,
  title={Rambo: Run-time packer analysis with multiple branch observation},
  author={Ugarte-Pedrero, Xabier and Balzarotti, Davide and Santos, Igor and Bringas, Pablo G},
  booktitle={International Conference on Detection of Intrusions and Malware, and Vulnerability Assessment},
  pages={186--206},
  year={2016},
  organization={Springer}
}

@misc{securityxploded_pespin_plugin,
  title        = {PESpin Plugin for ImpREC},
  author       = {{SecurityXploded}},
  howpublished = {\url{https://securityxploded.com/pespinplugin.php}},
  note         = {Accessed: 2026-03-26}
}

@misc{confuserex_github,
  author       = {{yck1509}},
  title        = {ConfuserEx},
  howpublished = {\url{https://github.com/yck1509/ConfuserEx}},
  year = {2015},
  note         = {GitHub repository, archived Jan. 27, 2019; accessed Mar. 26, 2026}
}

@misc{mpress,
    author = {MATCODE Software},
	title = {MPRESS is a free, high-performance executable packer for PE32/PE32+/.NET/MAC-DARWIN executable formats!
},
	howpublished = {\url{https://www.autohotkey.com/mpress/mpress_web.htm
}},
	note = {Accessed: 2026-03-26}
}

@misc{nsis,
  author       = {{Threat Intelligence Team}},
  title        = {{Revisiting the NSIS-based crypter
}},
  howpublished = {\url{https://www.threatdown.com/blog/revisiting-the-nsis-based-crypter/}},
  note         = {Accessed: July 17, 2025},
  year         = {2025}
}

@misc{pyinstaller_manual,
  author       = {{PyInstaller Development Team}},
  title        = {{PyInstaller Manual}},
  howpublished = {\url{https://pyinstaller.org/en/stable/}},
  note         = {Version 6.19.0, accessed: 2026-03-22}
}

@misc{molebox_website,
  author       = {Alexey Sudach{\'e}n},
  title        = {{Molebox Virtualization}},
  howpublished = {\url{https://sudachen.github.io/Molebox/}},
  note         = {Official website, accessed: 2026-03-22}
}

@article{bat2017entropy,
  title={Entropy analysis to classify unknown packing algorithms for malware detection},
  author={Bat-Erdene, Munkhbayar and Park, Hyundo and Li, Hongzhe and Lee, Heejo and Choi, Mahn-Soo},
  journal={International Journal of Information Security},
  volume={16},
  number={3},
  pages={227--248},
  year={2017},
  publisher={Springer}
}

@inproceedings{hai2017packer,
  title={Packer identification based on metadata signature},
  author={Hai, Nguyen Minh and Ogawa, Mizuhito and Tho, Quan Thanh},
  booktitle={Proceedings of the 7th software security, protection, and reverse engineering/software security and protection workshop},
  pages={1--11},
  year={2017}
}

@article{li2019consistently,
  title={A consistently-executing graph-based approach for malware packer identification},
  author={Li, Xingwei and Shan, Zheng and Liu, Fudong and Chen, Yihang and Hou, Yifan},
  journal={IEEE Access},
  volume={7},
  pages={51620--51629},
  year={2019},
  publisher={IEEE}
}

@inproceedings{saleh2017control,
  title={A control flow graph-based signature for packer identification},
  author={Saleh, Moustafa and Ratazzi, E Paul and Xu, Shouhuai},
  booktitle={MILCOM 2017-2017 IEEE Military Communications Conference (MILCOM)},
  pages={683--688},
  year={2017},
  organization={IEEE}
}

@article{zhang2018sensitive,
  title={Sensitive system calls based packed malware variants detection using principal component initialized MultiLayers neural networks},
  author={Zhang, Jixin and Zhang, Kehuan and Qin, Zheng and Yin, Hui and Wu, Qixin},
  journal={Cybersecurity},
  volume={1},
  number={1},
  pages={10},
  year={2018},
  publisher={Springer}
}

@inproceedings{sun2010pattern,
  title={Pattern recognition techniques for the classification of malware packers},
  author={Sun, Li and Versteeg, Steven and Bozta{\c{s}}, Serdar and Yann, Trevor},
  booktitle={Australasian Conference on Information Security and Privacy},
  pages={370--390},
  year={2010},
  organization={Springer}
}

@article{kancherla2016packer,
  title={Packer identification using Byte plot and Markov plot},
  author={Kancherla, Kesav and Donahue, John and Mukkamala, Srinivas},
  journal={Journal of Computer Virology and Hacking Techniques},
  volume={12},
  number={2},
  pages={101--111},
  year={2016},
  publisher={Springer}
}

@inproceedings{han2009packed,
  title={Packed PE file detection for malware forensics},
  author={Han, Seungwon and Lee, Keungi and Lee, Sangjin},
  booktitle={2009 2nd International Conference on Computer Science and Its Applications, CSA 2009},
  pages={5404211},
  year={2009}
}

@article{zhong2025unveiling,
  title={Unveiling Malware Visual Patterns: A Self-Analysis Perspective},
  author={Zhong, Fangtian and Hu, Qin and Jiang, Yili and Huang, Jiaqi and Cheng, Xiuzhen},
  journal={IEEE Transactions on Information Forensics and Security},
  year={2025},
  publisher={IEEE}
}

@article{zhong2024enhancing,
  title={Enhancing malware classification via self-similarity techniques},
  author={Zhong, Fangtian and Hu, Qin and Jiang, Yili and Huang, Jiaqi and Zhang, Cheng and Wu, Dinghao},
  journal={IEEE Transactions on Information Forensics and Security},
  year={2024},
  publisher={IEEE}
}

@misc{virusshare,
  author       = {{VirusShare}},
  title        = {VirusShare: A Repository of Malware Samples},
  howpublished = {\url{https://virusshare.com/}},
  note         = {Accessed Feb. 7, 2026}
}

@misc{Malpedia,
    author ={Fraunhofer FKIE},
	title = {Malpedia is a free service offered by Fraunhofer FKIE.
},
	howpublished = {\url{https://malpedia.caad.fkie.fraunhofer.de/}},
	note = {Accessed: 2026-03-26}
}

@misc{capev2,
  author       = {O'Reilly, Kevin and CAPEv2 contributors},
  title        = {{CAPEv2}: Malware Configuration and Payload Extraction},
  year         = {2026},
  howpublished = {\url{https://github.com/kevoreilly/CAPEv2}},
  note         = {GitHub repository, accessed May 25, 2026}
}

@inproceedings{vieira2010cohen,
  title={Cohen's kappa coefficient as a performance measure for feature selection},
  author={Vieira, Susana M and Kaymak, Uzay and Sousa, Jo{\~a}o MC},
  booktitle={International conference on fuzzy systems},
  pages={1--8},
  year={2010},
  organization={IEEE}
}

@misc{peid_github,
  author       = {{packing-box}},
  title        = {PEiD: Portable Executable Identifier},
  year         = {2026},
  howpublished = {\url{https://github.com/packing-box/peid}},
  note         = {GitHub repository. Accessed: Feb. 21, 2026}
}

@misc{readpe_github,
  author       = {{mentebinaria}},
  title        = {ReadPE: Portable Executable File Reader},
  year         = {2026},
  howpublished = {\url{https://github.com/mentebinaria/readpe}},
  note         = {GitHub repository. Accessed: Feb. 21, 2026}
}

@misc{die_github,
  author       = {{horsicq}},
  title        = {Detect It Easy (DiE)},
  year         = {2026},
  howpublished = {\url{https://github.com/horsicq/Detect-It-Easy}},
  note         = {GitHub repository. Accessed: Feb. 21, 2026}
}

@misc{kanxue_tools,
  author       = {{Kanxue}},
  title        = {Kanxue Tool},
  howpublished = {\url{https://tool.kanxue.com/}},
  note         = {Accessed Mar. 24, 2026}
}

@misc{awesome,
  author = {{packing-box}},
  title = {{Awesome Executable Packing}},
  year = {2026},
  url = {https://github.com/packing-box/awesome-executable-packing},
  note = {GitHub repository, accessed Mar. 24, 2026}
}

@misc{pyinstxtractor,
  author       = {{extremecoders-re}},
  title        = {PyInstaller Extractor},
  howpublished = {\url{https://github.com/extremecoders-re/pyinstxtractor}},
  note         = {GitHub repository, accessed Mar. 24, 2026}
}

@misc{pklite_data_unpacker,
  author       = {{Yt-trium}},
  title        = {PKLITE-1.12-data-unpacker},
  howpublished = {\url{https://github.com/Yt-trium/PKLITE-1.12-data-unpacker}},
  note         = {GitHub repository, accessed Mar. 24, 2026}
}

@misc{confuserex_unpacker,
  author       = {{XenocodeRCE}},
  title        = {ConfuserEx-Unpacker: A Dynamic Unpacker for ConfuserEx-Protected Binaries},
  howpublished = {\url{https://github.com/XenocodeRCE/ConfuserEx-Unpacker}},
  year         ={2017},
  note         = {GitHub repository, accessed Mar. 25, 2026}
}

@misc{themida_net_unpacker,
  author ={cg10036},
  title        = {Themida Unpacker for .NET},
  howpublished = {\url{https://github.com/cg10036/Themida-Unpacker-for-.NET}},
  note         = {GitHub repository, accessed Mar. 25, 2026}
}

@misc{quick_unpack,
  author       = {{fatrolls}},
  title        = {Quick-Unpack: A Fast Tool for Unpacking Executables},
  howpublished = {\url{https://github.com/fatrolls/Quick-Unpack}},
  note         = {GitHub repository, accessed Mar. 25, 2026}
}

@misc{fuu_unpack,
  author       = {{crackinglandia}},
  title        = {fuu: A Tool for Unpacking or Analyzing Packed Executables},
  howpublished = {\url{https://github.com/crackinglandia/fuu}},
  note         = {GitHub repository, accessed Mar. 25, 2026}
}

@misc{aldeid_winupack,
  author       = {{aldeid}},
  title        = {Category: Digital-Forensics/Computer-Forensics/Anti-Reverse-Engineering/Packers/WinUpack},
  year         = {2018},
  note         = {Last edited March 7, 2018. Accessed March 26, 2026},
  howpublished = {\url{https://www.aldeid.com/wiki/Category:Digital-Forensics/Computer-Forensics/Anti-Reverse-Engineering/Packers/WinUpack}}
}

@misc{mal_unpack,
  author       = {{hasherezade}},
  title        = {mal\_unpack: Dynamic Unpacker Based on PE-sieve},
  howpublished = {\url{https://github.com/hasherezade/mal_unpack}},
  note         = {GitHub repository, accessed Mar. 25, 2026}
}

@misc{unipacker,
  author       = {attilamester},
  title        = {UniPacker: Automatic and Platform-Independent Unpacker for Windows Binaries Based on Emulation},
  howpublished = {\url{https://github.com/unipacker/unipacker}},
  year         ={2025},
  note         = {GitHub repository, accessed Mar. 24, 2026}
}

@misc{orcastor_unpack,
  author       = {{orcastor}},
  title        = {unpack: An Unpacking Tool for PE Binaries},
  howpublished = {\url{https://github.com/orcastor/unpack}},
  note         = {GitHub repository, accessed Mar. 24, 2026}
}

\end{document}